\def\preprint{1}		

\ifdefined\preprint
  \documentclass[preprint,review,12pt]{elsarticle}
\fi
\ifdefined\wordcount
  \documentclass[final,3p,times,twocolumn]{elsarticle}
\fi
\ifdefined\final
  \documentclass[final,3p,times,twocolumn]{elsarticle}
\fi

\usepackage{lineno,hyperref}
\usepackage{placeins}
\usepackage{graphicx}
\usepackage{adjustbox}
\usepackage{subfigure}
\usepackage{color}
\usepackage{amsmath}
\usepackage{physics}
\usepackage{wrapfig}
\usepackage{chemformula}
\usepackage{placeins}
\usepackage{siunitx}
\usepackage{multirow}
\usepackage{booktabs}
\usepackage{tabularx} 
\journal{Combustion and Flame}

\biboptions{sort&compress}

\begin{document}

\begin{frontmatter}

\title{Three-dimensional numerical investigation of flashback in premixed hydrogen flames within perforated burners}


\author[1]{Filippo Fruzza}
\author[2]{Hongchao Chu}
\author[1]{Rachele Lamioni\corref{mycorrespondingauthor}}
\cortext[mycorrespondingauthor]{Corresponding author}
\ead{rachele.lamioni@unipi.it}
\address[1]{Department of Civil and Industrial Engineering, University of Pisa, Pisa 56122, Italy}
\author[3]{Temistocle Grenga}
\author[1]{Chiara Galletti}
\author[2]{Heinz Pitsch}
\address[2]{Institute for Combustion Technology, RWTH Aachen University, Aachen 52056, Germany}
\address[3]{Faculty of Engineering and Physical Sciences, University of Southampton, Southampton SO17 1BJ, UK}




\begin{abstract}
Predicting flashback represents a pivotal challenge in the development of innovative perforated burners for household appliances, especially for substituting natural gas with hydrogen as fuel. Most existing numerical studies have utilized two-dimensional (2D) simulations to investigate flashback in these burners, primarily to reduce computational costs. However, the inherent complexity of flashback phenomena suggests that 2D simulations may inadequately capture the flame dynamics, potentially leading to inaccurate estimations of flashback limits. In this study, three-dimensional (3D) simulations are employed to examine the impact of the actual slit shapes on the flashback velocities of hydrogen-premixed flames. Steady-state simulations are conducted to compute flashback velocities for three equivalence ratios ($\phi=0.6$, $0.8$, and $1.0$), investigating slits with fixed width $W$ and varying length $L$. Additionally, transient simulations are performed to investigate the flashback dynamics. The results are compared with those from 2D configurations to assess the reliability of the infinite slit approximation. For stable flames, 2D simulations underpredict the burner plate temperature compared to slits with lengths typical of practical devices but match the 3D results as $L\to\infty$. Conversely, flashback velocities are consistently underpredicted in 2D simulations compared to 3D simulations, even as $L\to\infty$. This is due to the critical role of the slit ends in flashback dynamics, where \textcolor{black}{favorable aerodynamics,} preferential diffusion, the Soret effect, and higher preheating due to a higher surface-to-volume ratio trigger the initiation of flashback in those regions. These findings underscore the necessity of employing 3D simulations to accurately estimate the flashback velocities in domestic perforated burners.
\end{abstract}

\begin{keyword}
Hydrogen \sep Flashback \sep Premixed flame \sep Perforated burner 
\end{keyword}
\end{frontmatter}

\section*{Novelty and significance statement}

\noindent This study presents a novel investigation into how finite slit lengths affect the critical flashback velocities in hydrogen-fueled perforated burners, using three-dimensional simulations. Our findings indicate that two-dimensional configurations, which are widely used in the literature, significantly underpredict flashback velocities because they fail to capture the crucial influence of slit ends. For the first time, we show that in slits of finite length and circular holes, \textcolor{black}{the combined effect of favorable aerodynamic conditions} and enhanced preheating, due to the increased surface area available for heat transfer, leads to higher flashback velocities compared to infinite-length slits. Additionally, we provide the first analysis of the temporal evolution of flashback dynamics in a realistic three-dimensional configuration, demonstrating that flashback initiation occurs at the slit ends. These insights are essential for the development of advanced numerical models that can inform the design of innovative perforated burners to prevent flashback effectively.

\section{Introduction}

As Europe strives to meet the ambitious targets set by the EU Green Deal, the role of green hydrogen has garnered significant attention~\cite{van_Renssen_2020,rastetter2013power}. One of its appealing applications is for the decarbonization of heating for residential and commercial buildings, especially in historical buildings, where the electrification is financially burdensome and technically complex~\cite{MCKENNA2018386_1,michalski2017hydrogen,SMITH202223071,DEVRIES2020114116,WU20228071,deVries2017impact}. These applications often rely on domestic end-user devices like condensing boilers equipped with premixed perforated burners. These burners typically consist of flat or cylindrical steel plates featuring patterns of tiny circular holes and slits which inject premixed fuel-air mixtures into combustion chambers, generating short flames that fit within the compact space between the burner and heat exchanger coils. The design of such burners has been developed and refined for decades to optimize their use with conventional fuels, especially natural gas~\cite{Najarnikoo2019,schiro2019experimental,EdacheriVeetil2018}. The introduction of hydrogen, either in its pure form or as a blend with natural gas, presents an opportunity for substantial \ch{CO2} emissions reduction but may require considerable changes in the burner design to ensure safety and efficiency. 

The unique challenges in fuelling hydrogen in domestic burners are related to the distinct physical properties of hydrogen compared to natural gas. Under stoichiometric conditions, the laminar flame speed of hydrogen can be about six times higher than natural gas~\cite{KONNOV2018197}. Furthermore, hydrogen exhibits a broader flammability range~\cite{SANCHEZ2014}. Lean hydrogen-air mixtures have a less-than-unity effective Lewis number and are, therefore, prone to thermo-diffusive instabilities~\cite{BERGER2022111935,chu2022PROCI,chu2023CNF}. As shown in several experimental and numerical works~\cite{FRUZZA2023,ANIELLO202233067,LAMIONI2023IJHE,Lamioni2022CTM,CUOCIchamber}, the use of weakly hydrogen-enriched mixtures (below 40\% by volume) in perforated burners does not cause drastic changes of the combustion characteristics, including pollutant emissions. However, when increasing the hydrogen content in the mixture, flashback phenomena emerge as one of the major issues related to the efficiency and the safety of hydrogen-fueled heating devices.

Flashback is the undesired upstream propagation of the flame into the premixing zone of the device. The flashback phenomenon represents a key issue also for many other applications, from laminar Bunsen flames to highly turbulent gas turbine combustors. For this reason, starting from the pioneering work by Lewis and von Elbe~\cite{LEWISVONELBE1943}, flashback limits have been studied for a number of fuels, applications, and operating conditions. Numerous distinct factors were identified to significantly influence the occurrence of flashback, such as preferential diffusion, stretch rate, Soret diffusion, heat losses to the walls, and wall temperatures~\cite{KEDIA20121055,VANCElewis,VANCEsoret,KURDYUMOV20071275,KIYMAZ202225022,KEDIA20151304,Altay2010}. These factors impact the balance between the flow velocity and the local flame speed, thereby determining both the occurrence of flashback and the flashback velocity, which is the critical gas velocity below which flashback occurs.

In light of the current emphasis on replacing natural gas with hydrogen in premixed perforated burners designed for condensing boilers, several experimental and numerical studies have been conducted in recent years to assess the flashback limits of \ch{H2}-air and \ch{H2}-\ch{CH4}-air flames. Aniello et al.~\cite{ANIELLO202233067} investigated experimentally the effect of \ch{H2} addition on blow-off and flashback limits, finding a strong correlation between the burner plate temperature and the flashback occurrence for high \ch{H2} contents. \textcolor{black}{Recent findings underscore the key role of heat losses to the burner walls in shaping flashback dynamics, particularly in very narrow slit configurations, where a distinct ``quenching failure" flashback regime arises due to the pronounced reduction in quenching distance at elevated wall and preheat temperatures~\cite{PERS2024quenching}.} In other related experimental investigations, Pers et al.~\cite{PERS2023autoignition,PERS2024autoignition} explored how the autoignition of the mixture impinging on the inner walls drives flashback when the burner plate reaches elevated temperatures, identifying this as a distinct ``wall-ignition" initiation regime compared to the ``hydrodynamic" regime.

On the numerical front, numerous studies have predominantly relied on two-dimensional (2D) simulations to explore flashback phenomena in premixed perforated burners. Vance et al.~\cite{VANCEcorrelation} utilized 2D simulations to investigate the flashback velocities of \ch{H2}-air flames, and proposed a novel Karlovitz number definition to correlate flashback conditions. Similarly, Fruzza et al.~\cite{FRUZZA2023} and Flores-Montoya et al.~\cite{FLORESMONTOYA2023113055} explored the flashback limits of \ch{H2}-enriched mixtures in multi-slit burners using 2D simulations. They identified two distinct flashback regimes: a symmetric bulk flashback and an asymmetric flashback, depending on the hydrogen content. Further advancing the field, Fruzza et al.~\cite{fruzzaUQ} applied stochastic sensitivity analysis methods in 2D simulations to quantify the relative effect of operating and geometric parameters on flashback. Additionally, Pers et al.~\cite{PERS2024113413} conducted 2D simulations to evaluate the impact of asymmetrical slit configurations on blow-off and flashback in \ch{CH4} and \ch{H2} laminar premixed burners.

\begin{figure}[!tb]
    \centering
    \includegraphics[width=0.9\textwidth]{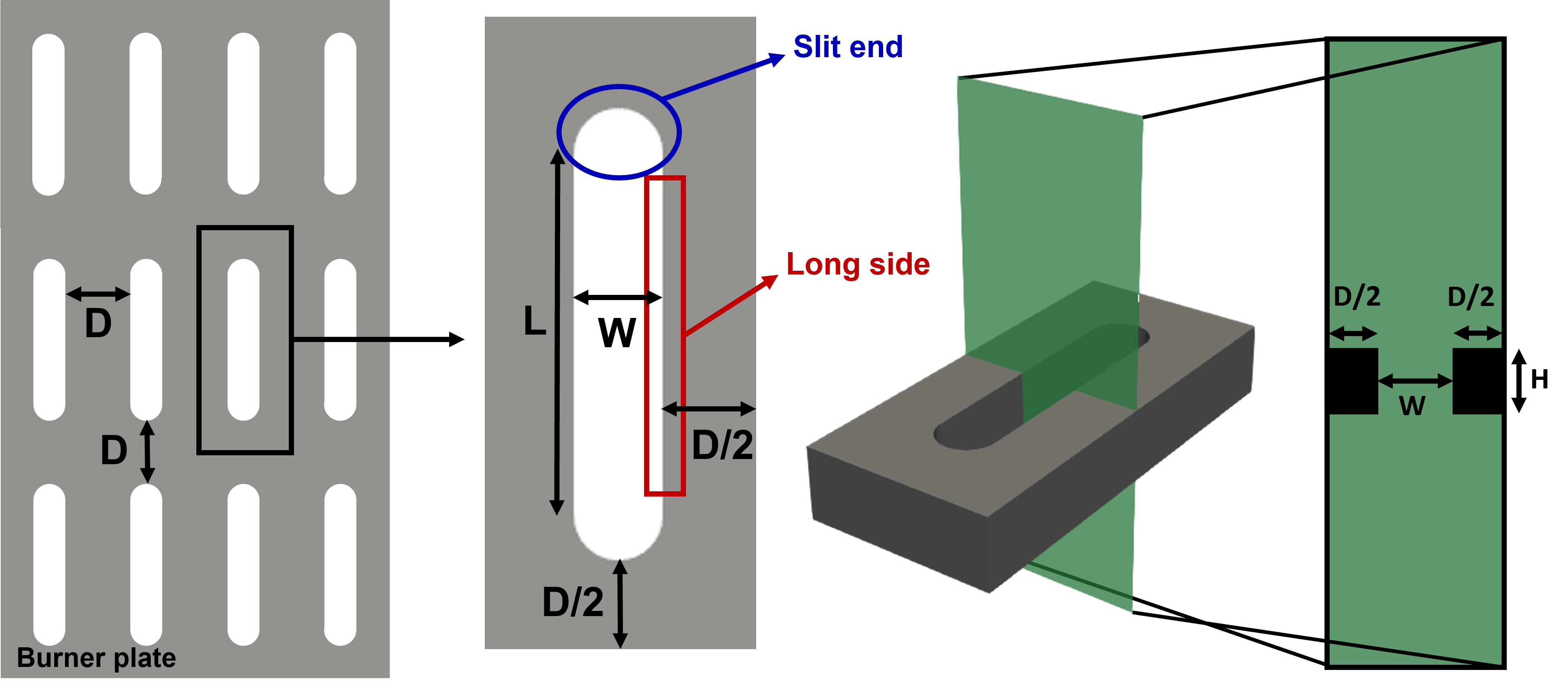}
    \caption{\footnotesize Left panel: slit pattern on the burner plate. Center panel: single slit geometry. Right panel: visualization of the 2D configuration.}
    \label{fig:plane_3D}
\end{figure}

All these numerical studies employed 2D configurations, which offer insights into the phenomenon and allow for an evaluation of the relevance of various parameters and physical mechanisms at a reduced computational cost. As depicted in Figure~\ref{fig:plane_3D}, this configuration represents a transversal section of an actual three-dimensional slit. The 2D geometry is defined by the slit width $W$, the spacing $D$ between adjacent slits, and the plate thickness $H$, neglecting the effects of the finite length $L$ (normal to the investigated plane) and the ends of the slit.
However, the validity of the infinite slit approximation, commonly assumed in 2D simulations, for practical slit geometries remains unassessed in the existing literature. In these simplified configurations, flashback is consistently predicted to initiate along the long edges of the slit. In contrast, real three-dimensional slits, with their finite lengths, may significantly alter flashback dynamics, potentially resulting in flashback velocities that differ markedly from those predicted by 2D configurations. Three-dimensional effects, particularly near the slit ends, are likely to influence the physical mechanisms driving flashback in premixed hydrogen flames, as suggested by our previous study~\cite{FRUZZA_PROCI}. In that work, we identified three primary mechanisms governing flashback behavior in perforated burners: preheating of the fresh gases, preferential diffusion, and the Soret effect. Through three-dimensional (3D) simulations, we demonstrated that the interaction of these mechanisms plays a critical role in determining flashback velocity, with a strong dependence on channel geometry. Notably, the Soret effect, when coupled with conjugate heat transfer, plays a dominant role in setting flashback limits through a self-reinforcing feedback mechanism that promotes flashback. Despite these findings, a direct comparison between 2D and 3D simulations has yet to be explored, and the influence of slit length on flashback velocity remains an open question.

In this study, we conduct numerical simulations in three- and two-dimensional configurations to investigate the influence of the finite length of the slit on the flashback velocities of \ch{H2}-air flames. Our research aims to answer three main questions: 
\begin{enumerate}
    \item How does slit length affect the shape of the flame and the heat transfer mechanism between the flame and the burner plate, and for which slit lengths is the 2D assumption valid?
    \item How does slit length influence the flashback velocity, and can a 2D configuration provide accurate estimates? 
    \item What is the dynamics of flashback in a realistic three-dimensional slit configuration?
\end{enumerate}
To explore these questions, we utilize both steady-state and transient three-dimensional simulations of flashback with a numerical domain representing a single slit within a perforated burner. The numerical model includes conjugate heat transfer between the gas phase and the burner plate. Steady-state simulations assess the impact of varying slit lengths on flame shape and burner plate temperature. The results are compared against those from 2D configurations to evaluate the reliability of the infinite slit approximation. A steady-state approach is also employed to compute the flashback velocity at three different equivalence ratios. Additionally, detailed transient simulations are conducted to investigate the differences in the temporal evolution of flashback between 2D and 3D slits.

\section{Configuration and numerical methods}

\begin{figure}[!tbph]
    \centering
    \includegraphics[width=\textwidth]{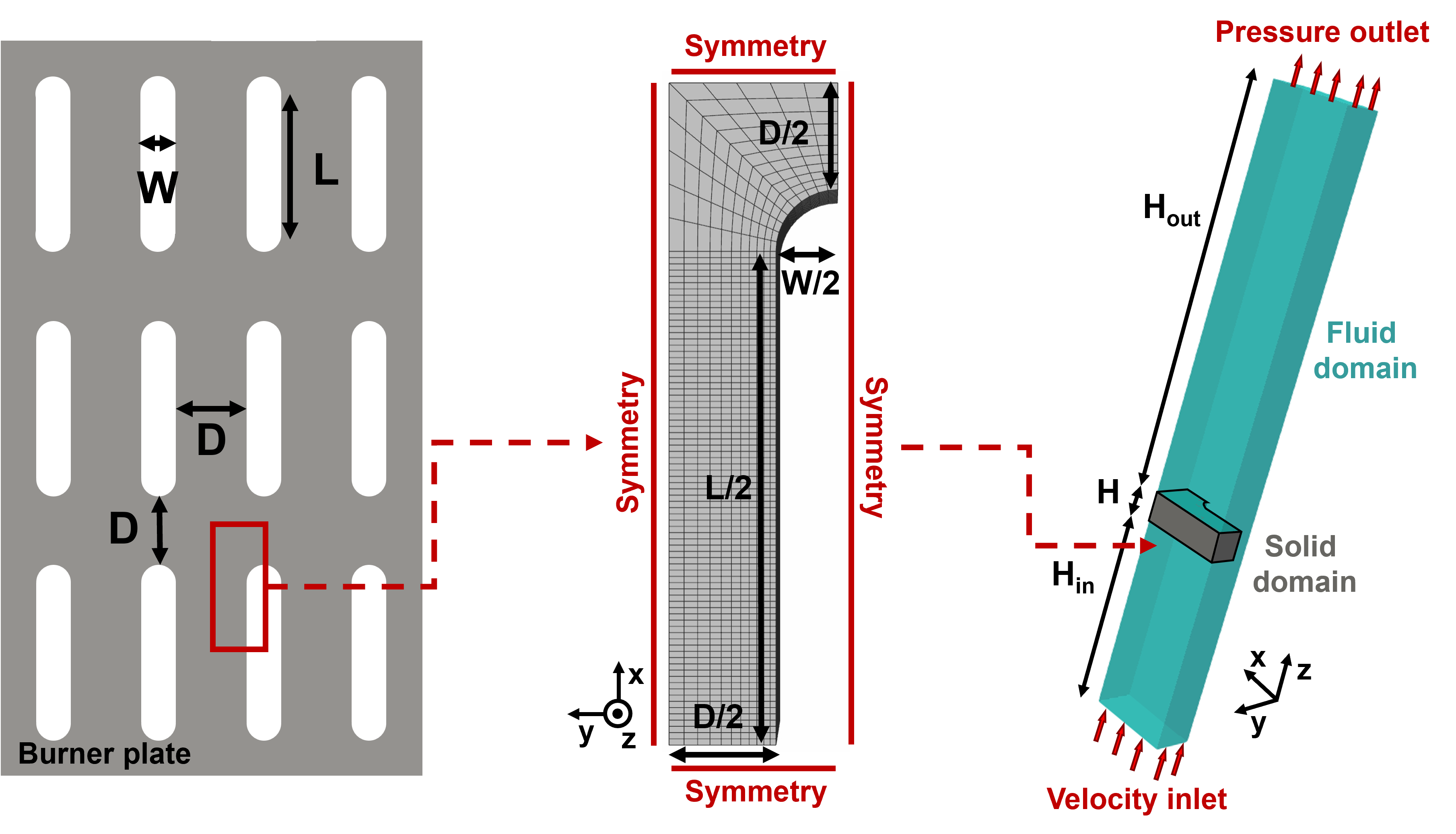}
    \caption{\footnotesize Left panel: Slit pattern on the burner plate, highlighting the section of the computational domain. Center panel: Slit geometry (boundary conditions in red) and solid plate mesh. Right panel: Computational domain (fluid zone in light blue, solid zone in grey, boundary conditions in red).}
    \label{fig:domain}
\end{figure}

In this study, we simulate a segment of the perforated burner plate typically used in domestic condensing boilers. 3D configurations representing arrays of holes or slits of different shapes and sizes are considered. The 3D configuration is shown in Figure~\ref{fig:domain} along with the computational domain. 
Due to the symmetries of the problem, the computational domain can be reduced to a quarter of the entire slit, with symmetry boundary conditions on the symmetry planes. The numerical fluid domain has a length of ${H_\mathrm{out}=\SI{8}{\mm}}$ downstream of the solid plate and a length of ${H_\mathrm{in}=\SI{4}{\mm}}$ upstream. $L=\SI{0}{\mm}$ corresponds to a circular hole of diameter $W$. The burner plate thickness is ${H=\SI{0.6}{\mm}}$ for all cases, and the slit width remains fixed at ${W=\SI{0.5}{\mm}}$. \textcolor{black}{A detailed investigation of the effects of varying slit width can be found in Fruzza et al.~\cite{FRUZZA_PROCI}.} For a given combination of $W$ and $L$, the distance between slits $D$ can be adjusted to fix the porosity of the burner, defined as $\psi=A_\mathrm{slit}/A_\mathrm{tot}$, where $A_\mathrm{slit}$ is the perforated area and $A_\mathrm{tot}$ is the sum of the plate and the perforated areas. We consider \ch{H2}-air mixtures at three equivalence ratios $\phi=0.6$, $0.8$, and $1.0$. Uniform velocities and a uniform temperature of ${T_u=\SI{300}{\K}}$ are set at the inlet, and a pressure outlet with ${p=\SI{1}{atm}}$ is imposed at the outlet. At the fluid-solid interface, a no-slip boundary condition is specified for the velocity, and zero-mass flux is imposed for the species equations. No thermal boundary conditions are necessary at this interface, as the fluid and solid domains are thermally coupled through the Conjugate Heat Transfer (CHT) approach, as described below.

In the fluid domain, the transport equations of mass, momentum, energy, and mass fractions of chemical species are given as:
\begin{gather}
    \frac{\partial\rho}{\partial t}+\boldsymbol{\nabla}\cdot\left(\rho \boldsymbol{v}\right)=0 \\
    \frac{\partial}{\partial t}\left(\rho \boldsymbol{v}\right)+\boldsymbol{\nabla}\cdot\left(\rho \boldsymbol{v}\boldsymbol{v}\right)=-\boldsymbol{\nabla}p+\boldsymbol{\nabla}\cdot\left(\boldsymbol{\Bar{\tau}}\right) \\
    \begin{split}
    & \frac{\partial}{\partial t}\left(\rho E\right)+\boldsymbol{\nabla}\cdot\left(\boldsymbol{v}\left(\rho E+p\right)\right)= \\
    & =\boldsymbol{\nabla}\cdot\left(k\boldsymbol{\nabla}T+\sum_{j=1}^N h_j \left(\sum_{k=1}^{N-1} \rho D_{m,jk}\boldsymbol{\nabla}Y_k+D_{T,j}\frac{\boldsymbol{\nabla}T}{T}\right)\right)-\sum_{j=1}^N h_j\dot{\omega}_j+S_{rad}
    \end{split} \\
    \frac{\partial}{\partial t}\left(\rho Y_i\right)+\boldsymbol{\nabla}\cdot\left(\rho\boldsymbol{v}Y_i\right)=\boldsymbol{\nabla}\cdot\left(\sum_{j=1}^{N-1} \rho D_{m,ij}\boldsymbol{\nabla}Y_j+D_{T,i}\frac{\boldsymbol{\nabla}T}{T}\right)+\dot{\omega}_i,
\end{gather}
\noindent where $\rho$ is the density, $\boldsymbol{v}$ is the velocity vector, $p$ is the pressure, and $\boldsymbol{\Bar{\tau}}$ is the stress tensor. The ideal gas law is applied as the equation of state. $T$ is the temperature. $N$, $h_i$, $Y_i$, and $\dot{\omega}_i$ are the number of species, the enthalpy, the mass fraction, and the net production rate of the $i$th species, respectively. The energy is defined as $E=\sum_{i=1}^N h_i Y_i-p/\rho+\abs{\boldsymbol{v}}^2/2$. $k$ is the mass-weighted thermal conductivity of the mixture, $D_{m,ij}$ are the generalized Fick’s law diffusion coefficients of the species $i$ in species $j$, and $D_{T,i}$ are the thermal diffusion coefficients of the $i$th species. Finally, $S_{rad}$ is the energy source associated with radiation. The equations are solved on a structured grid, with characteristic cell size in the reaction region of ${\Delta x = \SI{25}{\micro\metre}\simeq\delta_F/13}$, where $\delta_F$ is the 1D unstretched thermal flame thickness. \textcolor{black}{A grid independence study was performed, confirming that the selected resolution provides accurate results; the details of this study are provided in the Supplementary Material.} We employ detailed chemistry, using a reduced mechanism comprising 9 chemical species and 22 reactions. This reduced mechanism is derived from the Kee-58 skeletal mechanism~\cite{kee}. This choice ensures consistency with prior research involving \ch{CH4}-\ch{H2} mixtures up to 100\% \ch{H2}, where the same mechanism was used~\cite{FRUZZA2023,fruzzaUQ}. Full multicomponent diffusion is modeled through generalized Fick's law coefficients derived by the Maxwell-Stefan equations~\cite{fluent2022,taylorkrishna1993,Merk1959}. Soret diffusion is modeled using the following empirically-based composition-dependent expression provided by Kuo~\cite{kuo2012applications}:
\begin{equation}
    D_{T,i}=-2.59\times 10^{-7} T^{0.659}\left[\frac{M_{i}^{0.511}X_i}{\sum_{j=1}^N M_{j}^{0.511}X_j}-Y_i\right] \cdot \left[ \frac{\sum_{j=1}^N M_{j}^{0.511}X_j}{\sum_{j=1}^N M_{j}^{0.489}X_j} \right],
\end{equation}
\noindent where $M_i$, $X_i$, and $Y_i$ are the molar mass, molar fraction, and mass fraction of the species $i$, respectively. Radiation is modeled using the gray Discrete Ordinates (DO) method~\cite{MODEST2013541}, assuming the emissivity of the fluid-solid interface to be 0.85.

The burner plate is represented as a solid, embodying the properties of stainless steel commonly employed in such burners, including a density of ${\rho_s=\SI{7719}{\kg/\cubic\metre}}$, a specific heat capacity of ${c_{p,s}=\SI{461.3}{\J\per\kg\per\K}}$, and a thermal conductivity of ${k_s=\SI{22.54}{\W\per\metre\per\K}}$. Within the solid domain, the energy equation is solved as follows:
\begin{equation}
\frac{\partial}{\partial t}\left(\rho_s h_s\right)=\boldsymbol{\nabla}\cdot\left(k_s\boldsymbol{\nabla}T\right),
\end{equation}
where $h_s=\int_{T_{0}}^{T} c_{p,s} dT$ denotes the sensible enthalpy of the solid material. The CHT between the fluid and solid zones is modeled by using Fourier's Law to compute the heat flux on the fluid side at the fluid-solid interface, as ${q=k\left.\boldsymbol{n}\cdot\boldsymbol{\nabla}T\right|_{\mathrm{intf}}}$, where $\boldsymbol{n}$ represents the unit vector normal to the interface~\cite{fluent2022}. The walls are modeled as inert, a commonly used assumption in other similar works~\cite{KEDIA20121055,VANCEcorrelation,FRUZZA2023,FLORESMONTOYA2023113055} However, this approach may lead to elevated heat release rates at the wall in certain configurations due to the exclusion of heterogeneous catalytic reactions or surface chemistry~\cite{SANCHEZ2014,DENARDI2024}. Given the focus of this work on fluid-dynamic phenomena, this simplification is considered appropriate, though its potential impact on burner plate temperatures and flashback velocities is acknowledged.

\section{Solution methodology}\label{sec:solution}

In this study, two distinct solution approaches are employed to achieve various objectives. A steady-state approach is utilized to examine parametric variations across a wide range of simulations. Its computational efficiency makes it ideal for extensive parametric studies. It can provide information about the condition at which the flashback occurs, but it cannot be used to analyze the flashback dynamics. Conversely, a transient approach, despite being computationally demanding, is employed to study the flashback dynamics. A comparative analysis demonstrated that both approaches yield equivalent results in estimating burner temperatures and flashback velocities. \textcolor{black}{For further details, the reader is referred to the Supplementary Materials.}

\subsection{Steady-state approach}\label{sub:steady}

In the steady-state approach, a steady-state solver utilizing a pressure-based coupled algorithm~\cite{fluent2022} is employed along with a second-order scheme for spatial discretization. Initially, a stable flame solution is achieved by imposing a relatively high inlet velocity. Subsequently, the inlet velocity is systematically decreased until the steady-state solver fails to converge to a stable flame solution. This indicates the attainment of the critical inlet velocity for flashback. To accurately estimate the flashback inlet velocity, the minimum decrement of the inlet velocity is set to ${\Delta V_\mathrm{in}=\SI{0.01}{m/s}}$, where $V_\mathrm{in}$ is the uniform inlet velocity. Neglecting the density variations of the mixture due to the high burner plate temperatures, the cold-flow bulk velocity at the slit entry is defined as 
\begin{equation}\label{eq:defv}
    V_\mathrm{S}=\frac{A_\mathrm{tot}}{A_\mathrm{slit}}V_\mathrm{in}=\frac{1}{\psi}V_\mathrm{in},
\end{equation}
where $\psi$ is the porosity of the burner plate. \textcolor{black}{The choice of $V_\mathrm{S}$ as the cold-flow bulk velocity allows a direct and practical correlation to the burner’s thermal power, ensuring consistency across different geometries.} Following~\cite{VANCEcorrelation}, the flashback velocity $V_\mathrm{FB}$ is defined as the cold-flow bulk velocity at the slit entry when the flashback occurs, $V_\mathrm{FB}=\left.V_\mathrm{S}\right\vert_{\mathrm{FB}}$.

\subsection{Transient approach}\label{sub:transient}

In the transient approach, simulations are conducted on a domain encompassing the entire slit geometry, rather than a quarter of it, to capture potential asymmetries in the flashback dynamics. The PISO implicit algorithm is employed for this purpose~\cite{fluent2022}, with a second-order scheme for time discretization. The solution methodology remains consistent with the earlier description, maintaining a minimum velocity decrement of $\Delta V_\mathrm{in} = \SI{0.01}{\m/\s}$. A time step of $\Delta t = \SI{1}{\micro\s}$ is utilized within the fluid domain. However, the characteristic time scales of heat conduction within the solid domain significantly exceed those in the gas phase. As a result, a uniform time step across the entire computational domain would necessitate prohibitively long simulation times for the burner temperature to stabilize. To circumvent this, a larger time step for the solid zone, i.e., $\Delta t_s = 10^3 \Delta t$, is implemented. Similar methodologies have been previously applied in flashback simulations~\cite{FRUZZA2023, fruzzaUQ, FLORESMONTOYA2023113055}. \textcolor{black}{The chosen time step, validated in previous 2D studies~\cite{FRUZZA2023}, ensures accurate and stable results while maintaining computational efficiency.} It is important to note that while increasing the solid time step does not alter the estimated flashback velocities, it does affect the flashback dynamics. Specifically, while this method is effective for achieving steady-state solutions, it introduces unphysical flame oscillations during flashback due to the artificially accelerated heating of the solid. To mitigate this, once the flashback velocity is determined, the final simulation stage, corresponding to the flashback event, is conducted with a uniform time step of $\Delta t = \SI{1}{\micro\s}$ in both phases. This approach guarantees precise capture of flashback dynamics, devoid of any distortion due to unphysical heating effects within the solid domain.

\textcolor{black}{It is worth noting that, unlike some prior studies on \ch{CH4}-\ch{H2} mixtures, reporting auto-ignition at the wall due to elevated burner plate temperatures, our simulations did not exhibit such behavior. One possible reason is that the wall temperatures in our configurations remained below 1000 K, while auto-ignition typically occurs at higher temperatures as observed in experimental studies~\cite{PERS2023autoignition,PERS2024autoignition}. Additionally, while the regions between adjacent slits are included in our setup as the boundaries of the computational domain, the configuration is not explicitly designed to capture auto-ignition phenomena typically observed in these areas.}

\section{Results and discussion}

\textcolor{black}{In Section~\ref{sub:nr_flow}, we examine the aerodynamics and heat transfer mechanisms in a circular hole and a slit, focusing on the heat transfer from the burner plate to the flow. Non-reactive simulations with a heated burner plate are performed to analyze velocity profiles, gradients, and thermal interactions, providing insights into the potential influence of aerodynamic and thermal effects on flashback behavior.} In Section~\ref{sub:2dtemp}, we assess the accuracy of 2D simulations for predicting stable flames in 3D slits. The aim is to identify the critical length for which this configuration is valid and to examine its suitability for practical slits in actual burners. In Section~\ref{sub:fb}, flashback velocities are computed for slits of increasing length to examine the effect of the slit length. These results are also compared with the flashback velocities obtained from 2D simulations. Finally, Section~\ref{sub:dynamics} compares the flashback dynamics in 3D slits with that in 2D simulations, clarifying the underlying physics for the observations in previous sections.

\subsection{Aerodynamics and heat transfer in a circular hole and a slit}\label{sub:nr_flow}

\textcolor{black}{Understanding the interplay between local flow velocities, velocity gradients, heat transfer, and flame propagation speed is fundamental to the study of flashback phenomena. Velocity profiles are particularly critical, as regions of reduced velocity provide less resistance to flame propagation, increasing the likelihood of flashback~\cite{LEWISVONELBE1943,law_combphys}. Additionally, heat transfer from the burner plate to the flow can significantly influence the preheating of unburnt gases, potentially altering the velocity profiles as well as locally modifying the flame speed. All these factors are crucial for identifying potential flashback initiation points, where variations in the flow field and thermal conditions may promote or inhibit flame propagation locally.}

\textcolor{black}{To investigate the influence of aerodynamic and heat transfer effects in a circular hole and slit configuration, steady-state non-reactive simulations are performed to isolate these phenomena by neglecting the presence of a flame. We consider a circular hole with a diameter of $W = \SI{0.5}{\mm}$ and a slit with dimensions $L = \SI{2}{\mm}$ and $W = \SI{0.5}{\mm}$, both featuring a porosity of $\psi = 0.2$. For both cases, the inlet velocity is prescribed to achieve a velocity of $V_\mathrm{S} = \SI{3.5}{\m/\s}$ at the channel entry, with a uniform temperature of $\SI{300}{K}$. Isothermal boundary conditions are applied at the fluid-solid interface, with the burner plate maintained at $T_B = \SI{900}{K}$, representing a typical burner temperature associated with flashback condition.}

\textcolor{black}{The analysis focuses on a plane at $z = \SI{0.6}{\mm}$, located at the outlet of the hole/slit, where velocity fields, velocity gradients, and temperature distributions are examined. Figures~\ref{fig:nr_hole} and~\ref{fig:nr_slit} depict the $z$-velocity, $v_z$, and temperature fields for the circular hole and the slit, respectively. 
\begin{figure}[tbph]
    \centering
    \subfigure[]{\includegraphics[width=0.49\textwidth]{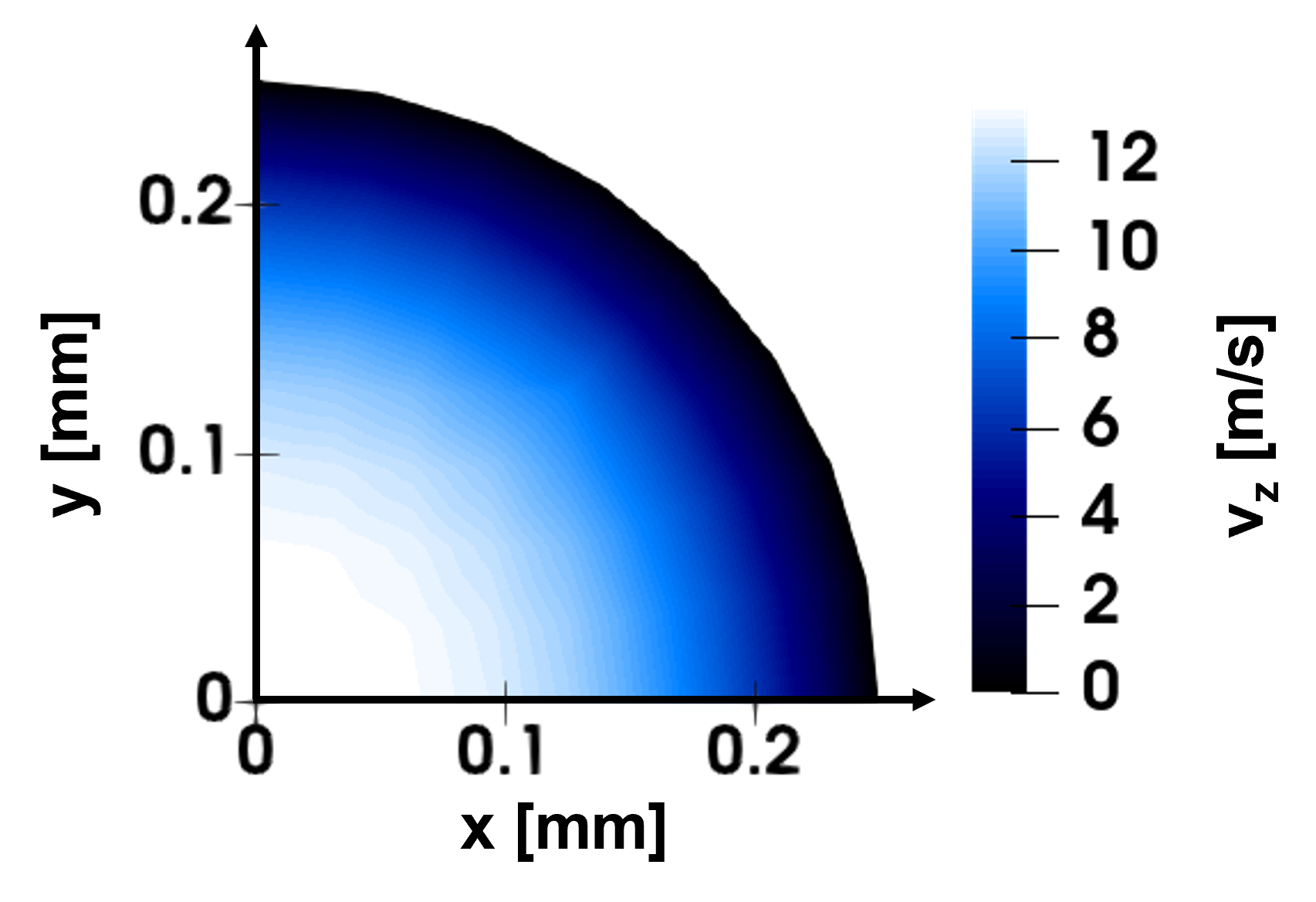}}
    \subfigure[]{\includegraphics[width=0.49\textwidth]{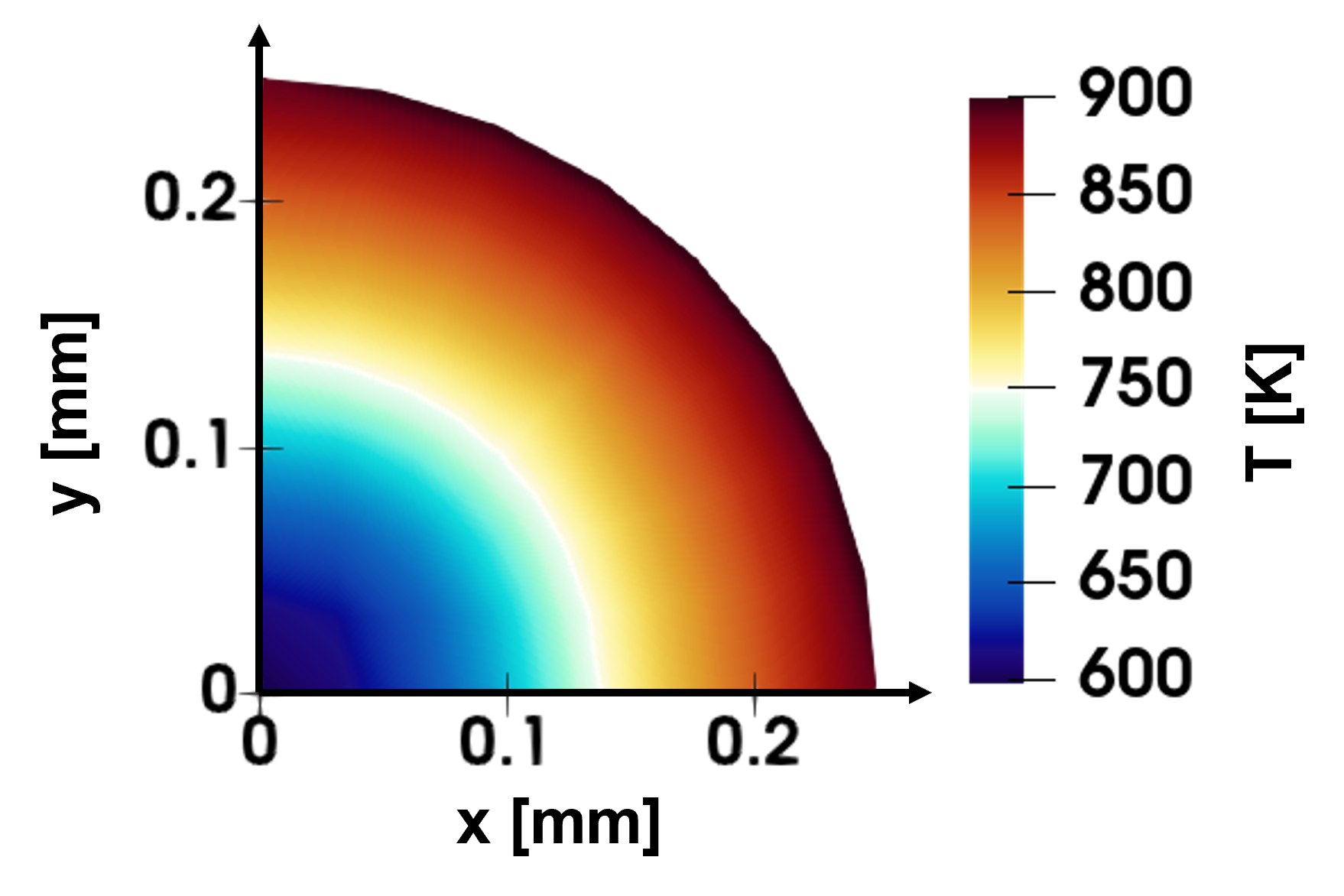}}
    \caption{\footnotesize Fields of $z$-velocity (a) and temperature (b) on a plane at $z = \SI{0.6}{\mm}$ for the circular hole.}
    \label{fig:nr_hole}
\end{figure}
\begin{figure}[tbph]
    \centering
    \subfigure[]{\includegraphics[width=0.8\textwidth]{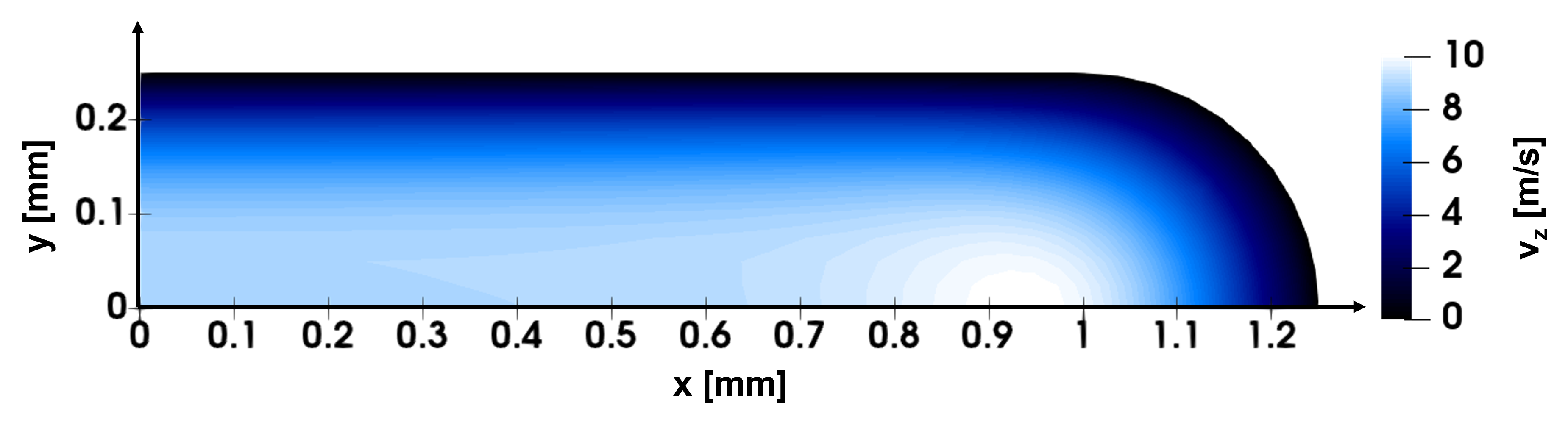}}
    \subfigure[]{\includegraphics[width=0.8\textwidth]{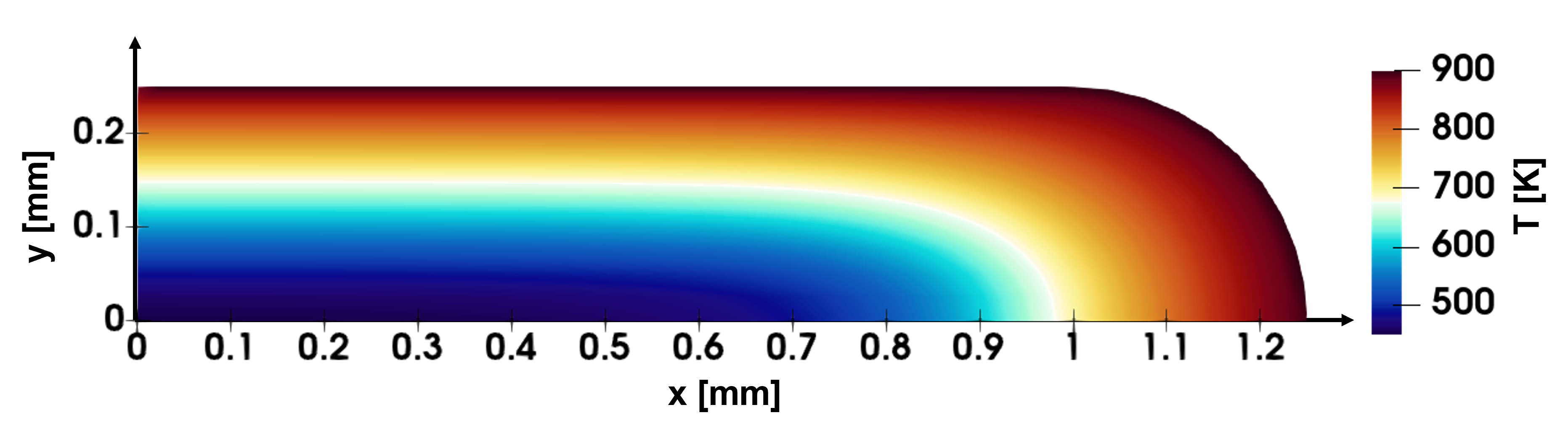}}
    \caption{\footnotesize Fields of $z$-velocity (a) and temperature (b) on a plane at $z = \SI{0.6}{\mm}$ for the slit.}
    \label{fig:nr_slit}
\end{figure}
The circular hole configuration exhibits a higher peak velocity than the slit, driven by the density reduction due to elevated temperatures. In the slit, the velocity reaches a relative peak near the slit ends, corresponding to an extended high-temperature region. The elevated temperatures in the circular hole and the slit ends, compared to the central slit region, result from enhanced heat transfer from the burner plate to the flow as it traverses the channel.} 

\textcolor{black}{Heat transfer efficiency in these configurations can be quantified using the surface-to-volume ratio, $\sigma$, defined as the ratio of the channel's internal surface area, where heat transfer occurs, to the volume of the channel through which the gas flows. For a circular hole with diameter $W$ and plate thickness $H$, the surface-to-volume ratio is 
\begin{equation}\label{sigmahole}
\sigma_{\mathrm{hole}} = \frac{\pi W \cdot H}{\pi (\frac{W}{2})^2 \cdot H} = \frac{4}{W}.
\end{equation}
\noindent In the slit configuration, the surface-to-volume ratio is non-uniform and varies across different regions of the channel. Consider a slit with length $L$, width $W$, and plate thickness $H$. For a segment of length $l$ located far from the slit ends, where the channel is rectangular, the local surface-to-volume ratio is 
\begin{equation}\label{sigmaslit}
    \sigma_\mathrm{slit} = \frac{2 \cdot l \cdot H}{W \cdot l \cdot H} = \frac{2}{W}.
\end{equation}
\noindent At the slit ends, which have a semicircular geometry with diameter $W$, the local surface-to-volume ratio equals that of the circular hole:
\begin{equation}\label{sigmaend}
\sigma_{\mathrm{end}} = \sigma_{\mathrm{hole}} = \frac{4}{W} = 2\sigma_\mathrm{slit}.
\end{equation}
\noindent The higher $\sigma$ at the slit ends and in the circular hole promotes enhanced heat transfer, leading to elevated exit temperatures and velocities. Consequently, the circular hole configuration exhibits higher temperatures and velocities overall, and similarly the slit shows localized enhancements near its ends.}

\begin{figure}[tbph]
    \centering
    \subfigure[]{\includegraphics[width=0.49\textwidth]{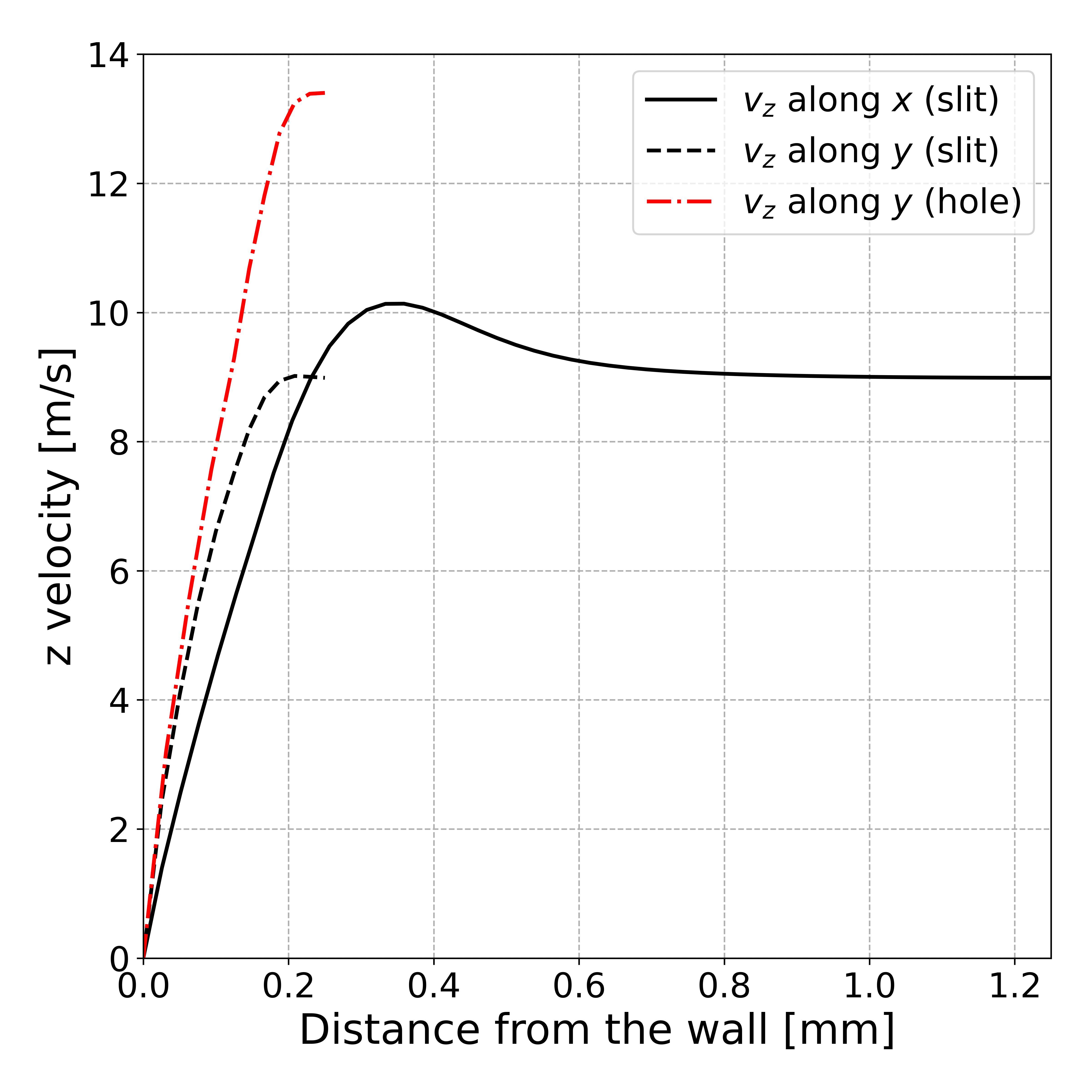}}
    \subfigure[]{\includegraphics[width=0.49\textwidth]{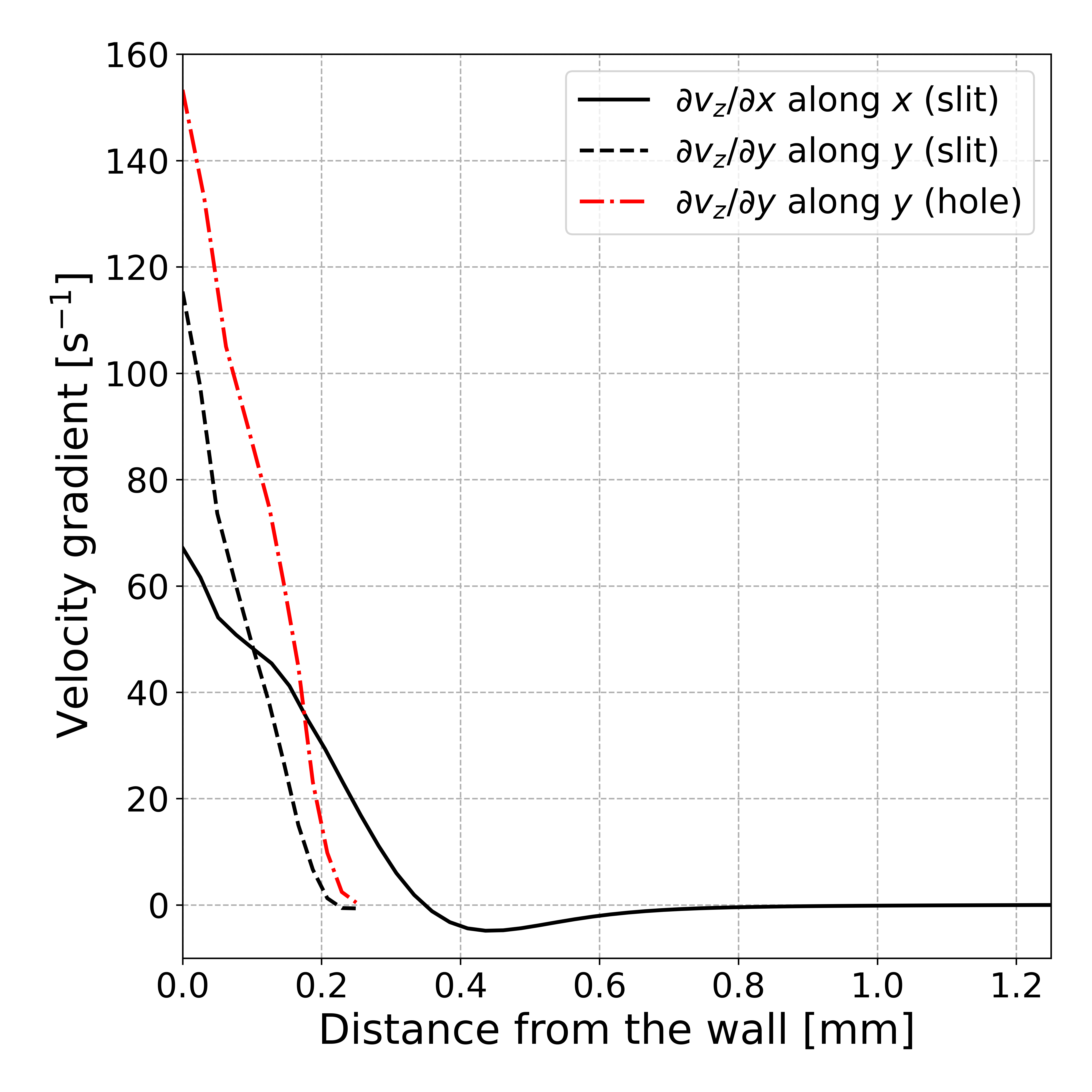}}
    \caption{\footnotesize Profiles of $z$-velocity (a) and velocity gradients (b) along the $y$ axis for the circular hole and along the $x$ (longitudinal) and $y$ (transverse) axes for the slit configuration, plotted as a function of the distance from the wall.}
    \label{fig:boundary_layer_profiles}
\end{figure}

\textcolor{black}{The boundary layer near the wall is the most critical zone for flashback
initiation. In this region, the reduced flow velocity, particularly at the location where the flame stabilizes, might allow the flame speed to exceed the local flow velocity, triggering flashback~\cite{LEWISVONELBE1943,KURDYUMOV20001883,VANCEcorrelation}. To investigate and compare boundary layer behavior across configurations, Figure~\ref{fig:boundary_layer_profiles}(a) shows the $z$-velocity ($v_z$) profiles along the $y$ axis for the circular hole configuration, and along both the $x$ (longitudinal) and $y$ (transverse) axes for the slit configuration at $z=\SI{0.6}{\mm}$. A schematic of these axes within the domain is provided in Figure~\ref{fig:domain} for reference. The velocity profiles are plotted as a function of the distance from the wall to enable direct comparison. Figure~\ref{fig:boundary_layer_profiles}(b) shows the corresponding velocity gradient profiles, with $\partial v_z/\partial y$ for the circular hole and both $\partial v_z/\partial x$ (longitudinal) and $\partial v_z/\partial y$ (transverse) for the slit configuration.}
\noindent\textcolor{black}{As previously noted, the circular hole exhibits a significantly higher velocity compared to the slit, resulting in a higher velocity gradient. Within the slit, the boundary layers in the transverse and longitudinal directions show distinct behavior due to the asymmetrical geometry, where $L \gg W$. A peak in $v_z$ is observed along the longitudinal axis because of the elevated temperature in that region. However, the velocity increases more sharply in the transverse direction, as evidenced by the velocity gradient, whose maximum is nearly double that observed along the longitudinal direction.}

\textcolor{black}{The results reveal significant differences between the circular hole and slit configurations. The circular hole exhibits higher temperatures and steeper velocity gradients due to its confined geometry and high surface-to-volume ratio, which enhances heat transfer efficiency. In contrast, the slit configuration shows more complex behavior with pronounced internal variations. Within the slit, the central region and the ends display distinct characteristics. The central region is marked by high velocity gradients in the transverse direction, demonstrating higher aerodynamic resistance to flashback. In comparison, the slit ends exhibit velocity peaks due to intensified heating in these localized zones, but the velocity gradients near the slit ends are lower than those in the central region. This combination of high temperatures and low velocity gradients makes the slit ends critical zones for flashback initiation. These findings underscore the importance of considering both global and localized effects when analyzing flow configurations.  However, aerodynamics alone does not fully determine flashback, as it is governed by the interaction between flow velocity and flame speed. The latter can vary locally due to factors such as preheating of unburnt gases and local fuel enrichment. The next sections will explore the flame morphology and flashback process in greater detail.}

\subsection{Assessment of the validity of the infinite slit approximation}\label{sub:2dtemp}

Previous numerical studies investigating flashback in multi-slit burners have exclusively utilized 2D configurations~\cite{VANCEcorrelation,FRUZZA2023,FLORESMONTOYA2023113055,fruzzaUQ}. In a 2D simulation representing a slit array, the simulation domain models a cross-sectional profile of the slit, as illustrated in Figure~\ref{fig:plane_3D}. The geometry of this profile is defined by the slit width $W$, the spacing $D$ between adjacent slits, and the plate thickness $H$. In 2D, the porosity is $\psi=W/(W+D)$. Such a configuration, being invariant with respect to the actual slit length, assumes the slit to be infinitely long, thus omitting the ends of the slit. This section aims to determine the minimal slit length at which 2D simulations can accurately approximate a 3D slit configuration. To this end, we perform a comparative analysis of stable flames derived from steady-state simulations for 3D configurations with varying slit lengths, from $\SI{1}{\mm}$ to $\SI{80}{\mm}$, and a corresponding 2D configuration. The geometric parameters are standardized so that all 3D geometries have identical cross-sectional profiles, allowing for direct comparison with the 2D domain. Note that in the 3D configurations, the porosity varies for different slit lengths, approaching the 2D value as $L\to\infty$. To maintain consistency across comparisons, the inlet velocity is adjusted to ensure the same velocity at the slit entry for all simulations, set as $V_\mathrm{S}=\SI{4}{\m/\s}$. The equivalence ratio of the mixture is set at $\phi=0.6$. The specific geometric and operating parameters utilized are listed in Table~\ref{tab:2dtemp}.

\begin{table}[htbp]
\centering
\caption{Overview of the simulation cases described in Section~\ref{sub:2dtemp}.}
\label{tab:2dtemp}
{\small\begin{tabular}{@{}ccccccc@{}}\toprule
\multicolumn{5}{c}{\textbf{Geometry}}&\multicolumn{2}{c}{\textbf{Op. Parameters}}\\ \midrule
\textbf{Config.}&$\boldsymbol{L}$ \textbf{[mm]}&$\boldsymbol{W}$ \textbf{[mm]}&$\boldsymbol{D}$ \textbf{[mm]}&\textbf{$\boldsymbol{H}$} \textbf{[mm]}&$\boldsymbol{V_\mathrm{S}}$ \textbf{[m/s]}&$\boldsymbol{\phi}$ \textbf{[-]}\\ \midrule
2D&-&0.5&1&0.6&4.0&0.6\\
\multirow{2}{*}{3D}&\multirow{2}{*}{\shortstack{1, 2, 5, 10, 20,\\ 35, 50, 80}}&\multirow{2}{*}{0.5}&\multirow{2}{*}{1}&\multirow{2}{*}{0.6}&\multirow{2}{*}{4.0}&\multirow{2}{*}{0.6}\\
&&&&&&\\ \bottomrule
\end{tabular}}
\end{table}

\begin{figure}[!tb]
    \centering
    \includegraphics[width=1.0\textwidth]{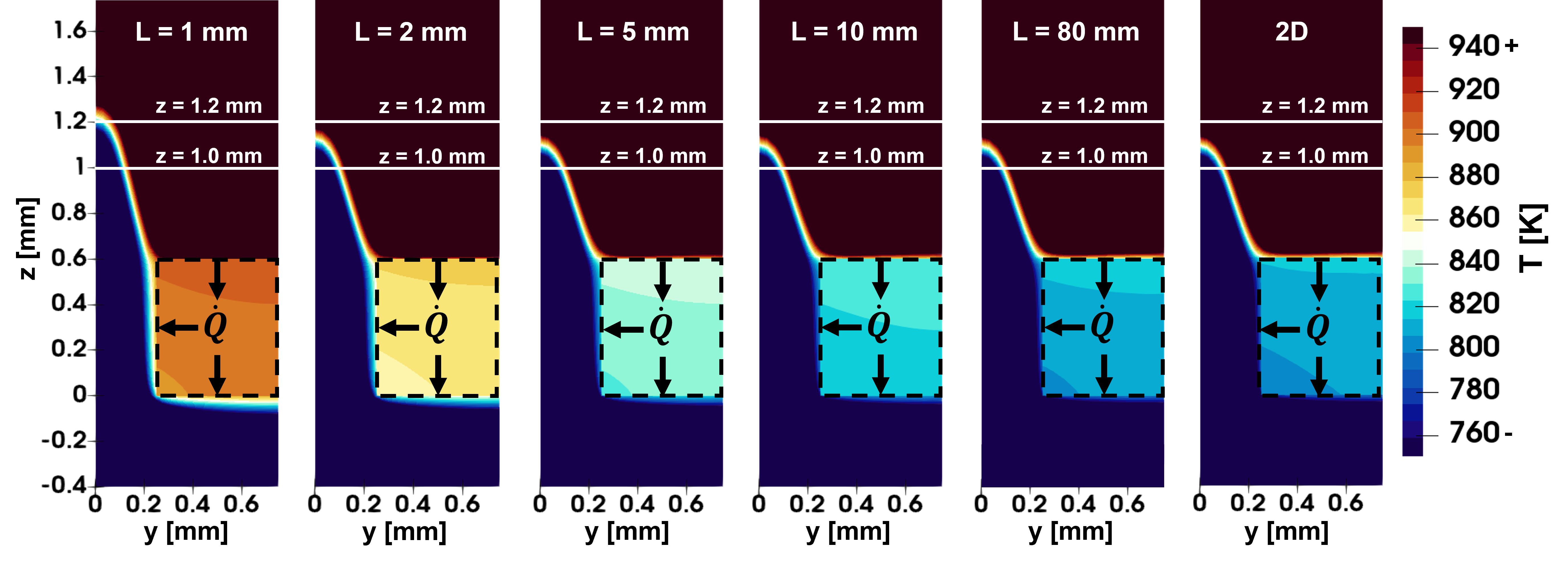}
    \caption{\footnotesize Temperature fields on the transversal plane at the center of the slit for 3D configurations with slit lengths of $1, 2, 5, 10, 80\ \SI{}{\mm}$, compared to the 2D configuration. The color bar is calibrated to highlight changes in burner plate temperature.}
    \label{fig:stretch}
\end{figure}

\begin{figure}[tbph]
    \centering
    \includegraphics[width=0.7\textwidth]{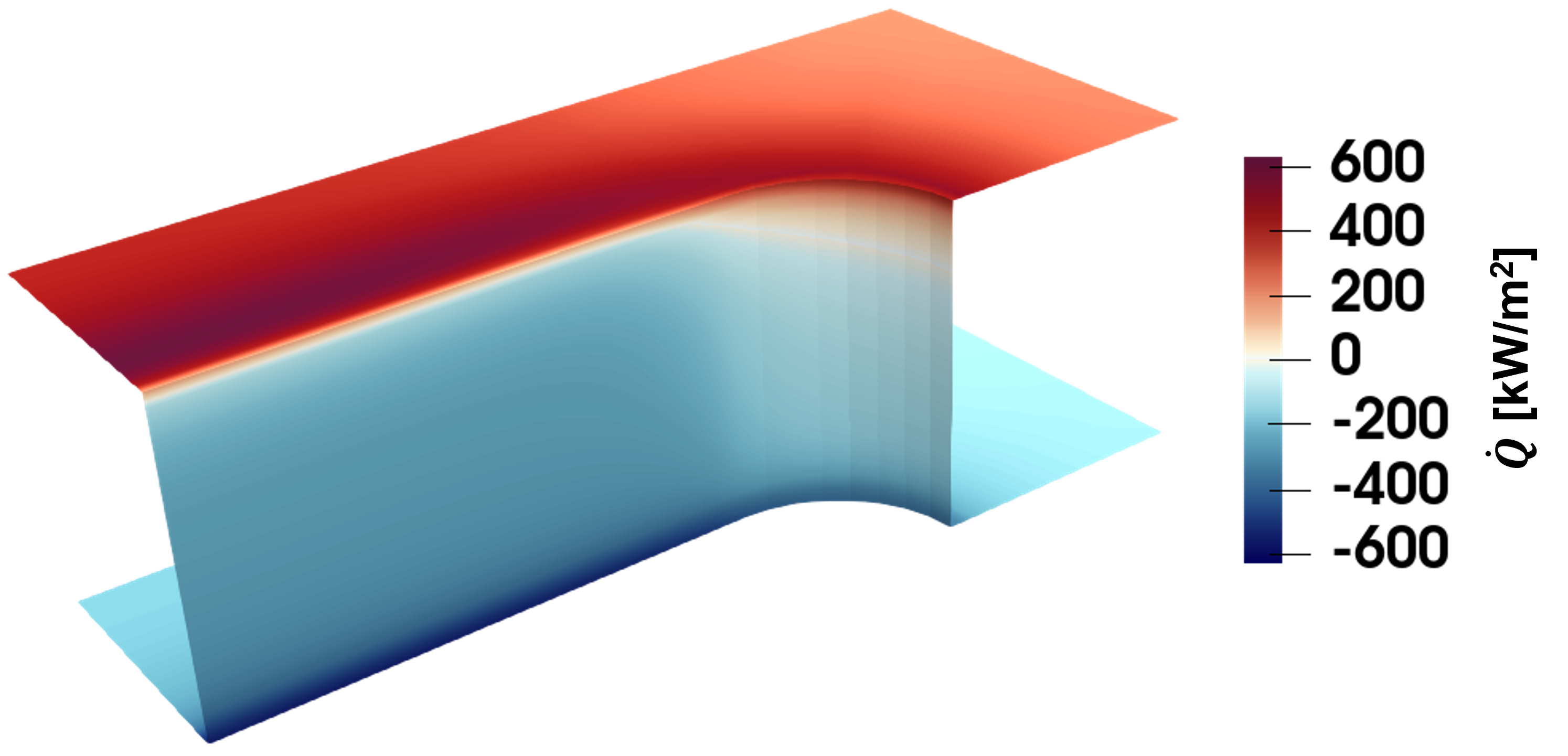}
    \caption{\footnotesize Heat flux distribution at the fluid-solid interface for the case ${L=\SI{2}{\mm}}$. Positive values indicate heat transfer from the gas phase to the solid phase.}
    \label{fig:heat_flux}
\end{figure}

To illustrate the impact of slit length and facilitate comparison with the 2D simulation, Figure~\ref{fig:stretch} displays temperature fields on the transversal section of the slit. The figure includes profiles from 3D configurations with varying slit lengths $L=1, 2, 5, 10, 80\ \SI{}{\mm}$, as well as from the 2D configuration. The heat flux is indicated as $\Dot{Q}$, and the black arrows indicate the direction of the heat flow. The color bar is calibrated to highlight variations in burner plate temperature and to distinctly display the flame front. This emphasis on burner plate temperature is crucial, as it significantly influences the flashback velocity~\cite{FRUZZA2023}.
\begin{figure}[tbph]
    \centering
    \includegraphics[width=0.7\textwidth]{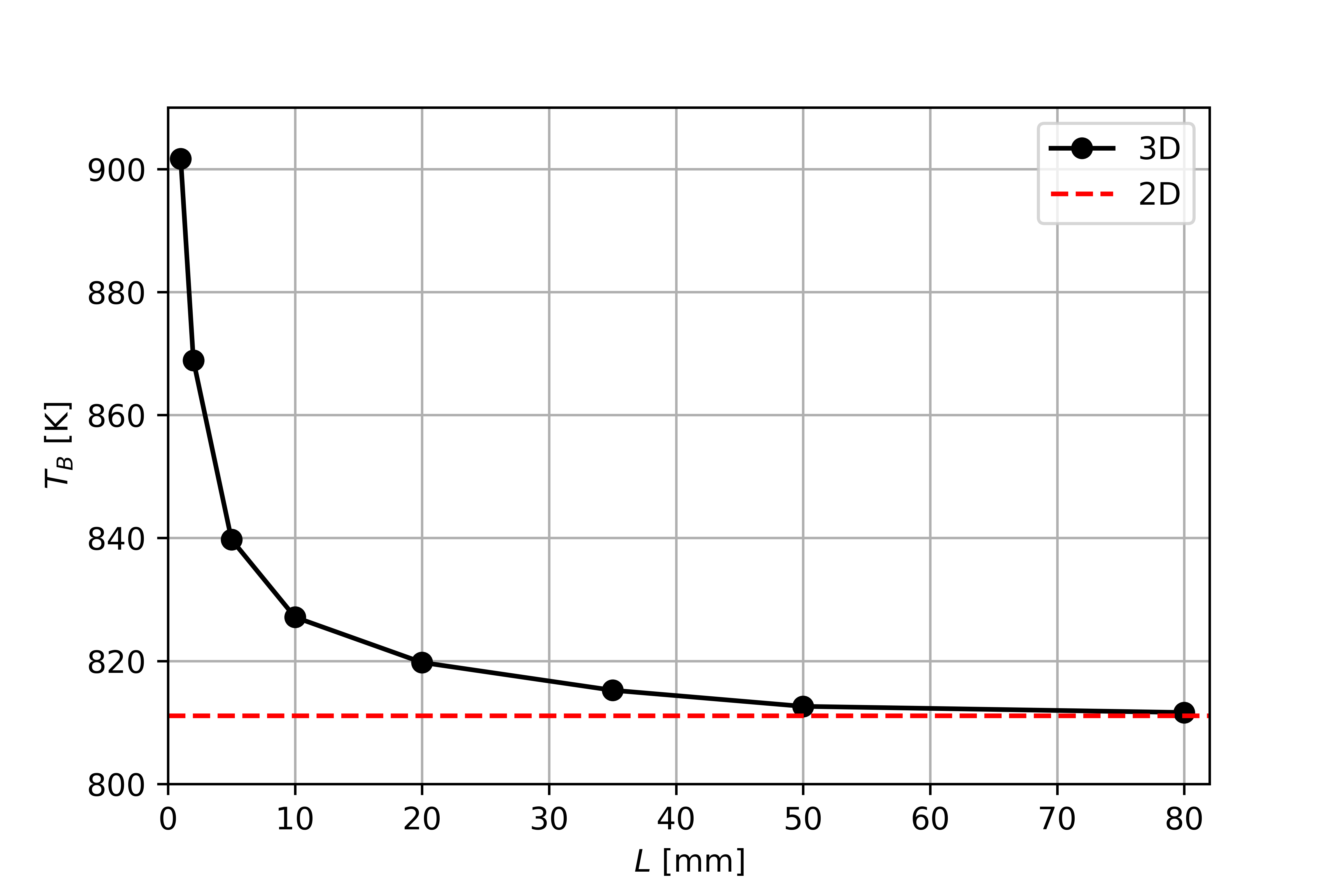}
    \caption{\footnotesize Average burner plate temperature as a function of the slit length for ${W=\SI{0.5}{\mm}}$ and ${D=\SI{1}{\mm}}$. The red dashed line represents the 2D result.}
    \label{fig:2D_3D_comparison}
\end{figure}
For ${L>\SI{2}{\mm}}$, the flame tip position is independent of slit length and is well predicted by the 2D simulation. Conversely, notable discrepancies are observed in the burner plate temperatures. \textcolor{black}{Interestingly, for $L=\SI{1}{\mm}$, the flame front appears higher in the two-dimensional cross-sectional view shown in Figure~\ref{fig:stretch}, which may seem counterintuitive given the higher burner temperature. This apparent inconsistency arises because the figure represents only a 2D slice of the domain. A full 3D analysis (not shown here for brevity) reveals that, for $L=\SI{1}{\mm}$, most of the flame front is actually positioned lower compared to longer slits, particularly at the slit ends.} To further illustrate how the burner temperature depends on slit length, Figure~\ref{fig:2D_3D_comparison} plots its volume-averaged value in the solid domain as a function of the slit length, including results from the 2D simulation for comparison. The 2D simulation forecasts a burner temperature of ${T_B=\SI{812}{\K}}$. In contrast, 3D simulations indicate a higher burner temperature of ${T_B=\SI{902}{\K}}$ for ${L=\SI{1}{\mm}}$, which gradually decreases to ${T_B=\SI{827}{\K}}$ at ${L=\SI{10}{\mm}}$, and eventually aligns with the 2D result for larger $L$. The high temperatures in short slits can be attributed to the more pronounced influence of the slit ends, where additional heat transfer occurs from the flame and burnt gases to the burner plate. Figure~\ref{fig:heat_flux} shows the heat flux profile at the fluid-solid interface for the case ${L=\SI{2}{\mm}}$. Here, the heat lost by the flame and burnt gases at the top face of the burner plate is reintroduced to the flow at the inner and bottom faces of the burner plate.
In shorter slits, the heat transferred to the burner plate at the ends represents a considerable fraction. As the slit length increases, the influence of the slit ends diminishes, aligning more closely with the conditions modeled in 2D simulations where these regions are neglected. This difference in burner plate temperatures between short slits and the 2D configuration is significant, particularly when estimating flashback velocity, which is highly sensitive to the burner plate temperature. According to Fruzza et al.~\cite{FRUZZA2023}, a temperature difference of ${\SI{90}{\K}}$ can lead to an underestimation of the flashback velocity by 30-40\%. While the 2D simulations effectively approximate the behavior of 3D slits when they are longer than $\SI{50}{\mm}$, it is crucial to note that slits in practical condensing boiler burners are substantially shorter, typically ranging from $\SI{0}{\mm}$ to $\SI{5}{\mm}$. This discrepancy indicates that 2D simulations fail to capture the true behavior of actual burners, emphasizing the need for 3D simulations of practical devices.

\subsection{Effect of the slit length on the flashback velocity}\label{sub:fb}

In this section, we examine the influence of slit length on flashback velocities and conduct a comparative analysis with results from 2D simulations. Flashback velocities are computed using the steady-state approach outlined in Section~\ref{sub:steady} for three equivalence ratios, i.e., $\phi=0.6$, $0.8$, and $1.0$. We consider slits of increasing lengths ranging from ${L = \SI{0}{\mm}}$ to ${\SI{8}{\mm}}$, with ${L = \SI{0}{\mm}}$ representing a circular hole. Longer slits are excluded as they are not practically relevant for real burners. To ensure comparability across cases, a fixed porosity of $\psi=0.2$ is maintained by adjusting the spacing between slits, $D$. Fixing the porosity is crucial as it ensures that the relationship given in Equation~\ref{eq:defv} between the inlet velocity, $V_\mathrm{in}$, and the cold-flow velocity at the slit entry, $V_\mathrm{S}$, remains consistent across all cases. Consequently, a given $V_\mathrm{S}$ corresponds to the same thermal power for all configurations, mirroring the performance of various ideal burners operating under identical conditions. Additionally, the results include the flashback velocities computed for a 2D configuration representing an infinite slit, using the same width and porosity as those imposed in the 3D cases. Detailed descriptions of the specific geometric and operating parameters used are provided in Table~\ref{tab:fb}.
\begin{table}[htbp]
\centering
\caption{Overview of the simulation cases described in Section~\ref{sub:fb}.}
\label{tab:fb}
{\small\begin{tabular}{@{}ccccccc@{}}\toprule
\multicolumn{5}{c}{\textbf{Geometry}}&\multicolumn{1}{c}{\textbf{Op. parameters}}\\ \midrule
\textbf{Config.}&$\boldsymbol{L}$ \textbf{[mm]}&$\boldsymbol{W}$ \textbf{[mm]}&$\boldsymbol{\psi}$ \textbf{[-]}&\textbf{$\boldsymbol{H}$} \textbf{[mm]}&$\boldsymbol{\phi}$ \textbf{[-]}\\ \midrule
2D&-&0.5&0.2&0.6&0.6, 0.8, 1.0\\
\multirow{2}{*}{3D}&\multirow{2}{*}{\shortstack{0, 0.5, 1, 2, 4, 8}}&\multirow{2}{*}{0.5}&\multirow{2}{*}{0.2}&\multirow{2}{*}{0.6}&\multirow{2}{*}{0.6, 0.8, 1.0}\\
&&&&&&\\ \bottomrule
\end{tabular}}
\end{table}
 
\begin{figure}[tbph]
    \centering
    \subfigure[]{\includegraphics[width=0.45\textwidth]{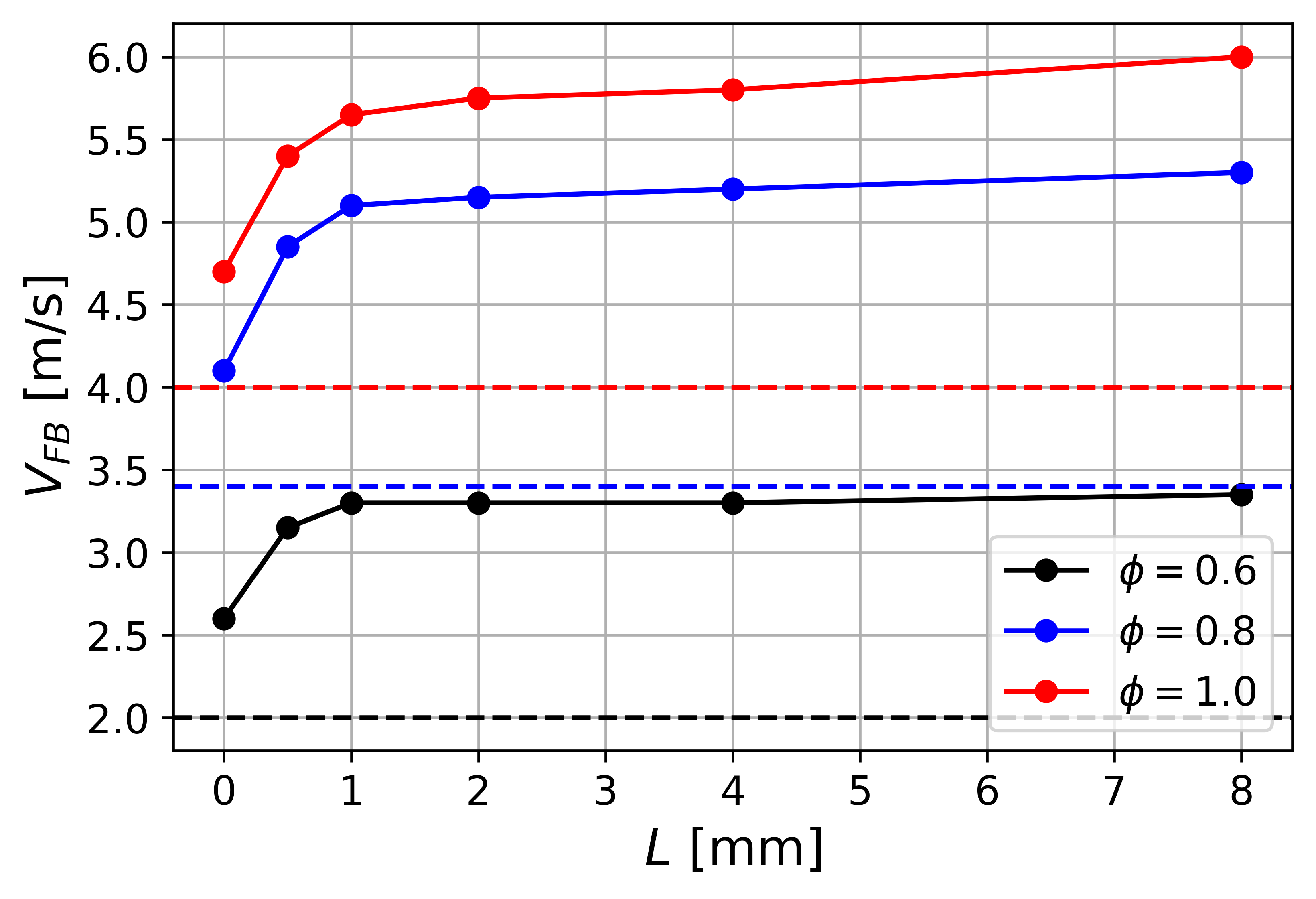}}
    \subfigure[]{\includegraphics[width=0.45\textwidth]{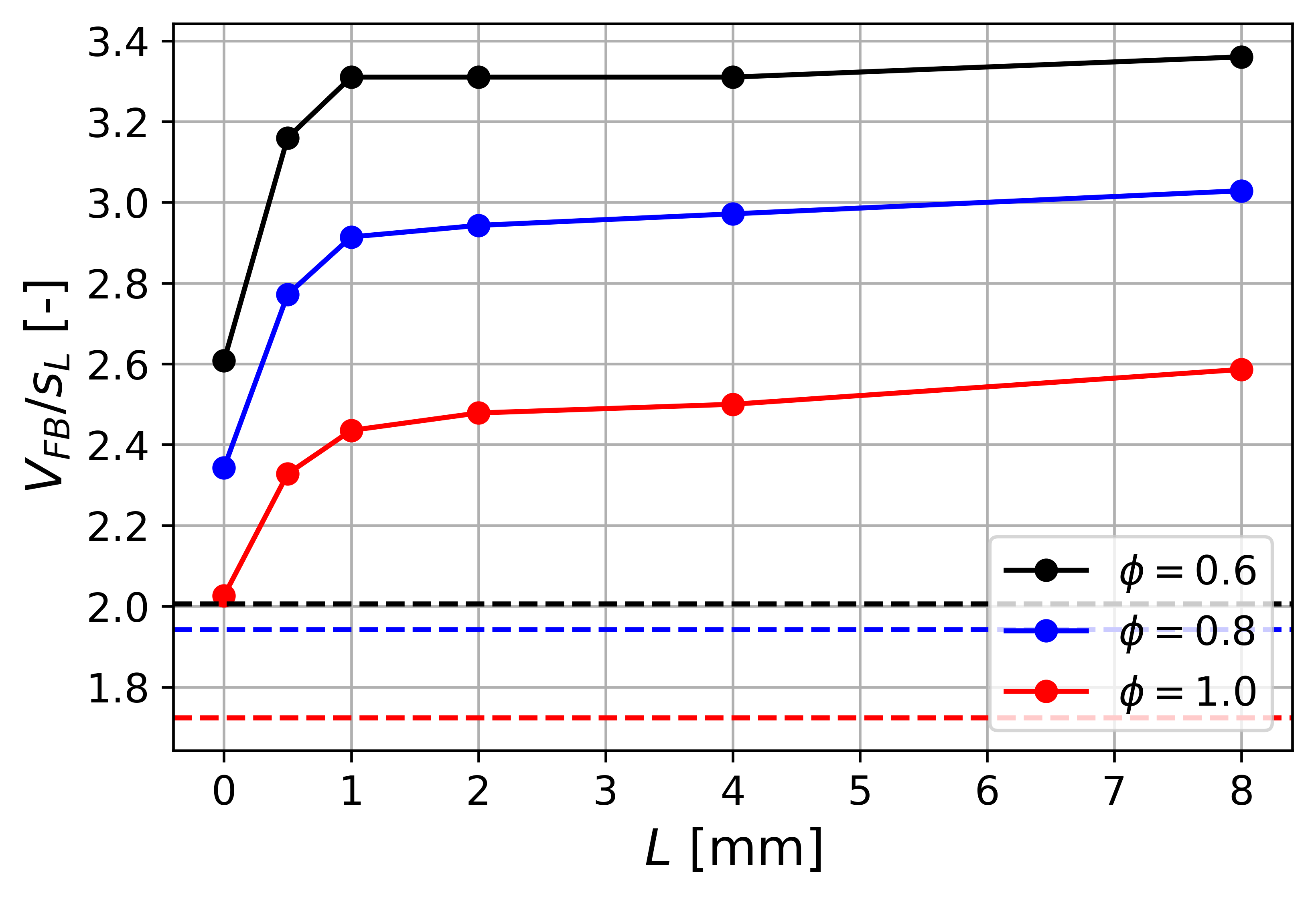}}
    \caption{\footnotesize Flashback velocity as a function of the slit length for various equivalence ratios: (a) non-normalized values and (b) normalized values with respect to $s_L$. 2D results are indicated by dashed lines. For all cases, $W=\SI{0.5}{\mm}$ and $\psi=0.2$.}
    \label{fig:Vfb_L}
\end{figure}

In Figure~\ref{fig:Vfb_L}, we plot the flashback velocity, $V_\mathrm{FB}$, against slit length, $L$, for three equivalence ratios. Panel (a) presents the values of the flashback velocity, while panel (b) shows these values normalized by the 1D unstretched laminar flame speed, $s_L$, at the nominal conditions.
As expected, both 2D and 3D simulations show that $V_\mathrm{FB}$ increases with the equivalence ratio, reflecting the higher laminar flame speeds associated with richer mixtures. In contrast, the normalized flashback velocity, $V_\mathrm{FB}/s_L$, which more accurately reflects flashback propensity by removing the trivial dependence on laminar flame speed, exhibits a decreasing trend in both settings. This relationship is expected because flashback in hydrogen flames is strongly influenced by preferential diffusion effects, with leaner mixtures generally showing a higher propensity for flashback~\cite{VANCEcorrelation,fruzzaUQ}. The flashback velocity shows a rapid increase from $L=\SI{0}{\mm}$ to $L=\SI{1}{\mm}$. However, for lengths greater than $L=\SI{1}{\mm}$, the influence of slit length on flashback velocity significantly diminishes. This finding indicates a distinct behavior of circular holes and very short slits in contrast to longer slits, which exhibit relatively uniform behavior regardless of their length. Notably, unlike the burner plate temperatures shown in Figure~\ref{fig:2D_3D_comparison}, where the 3D results eventually align with the 2D values for larger slit lengths, the 3D results for flashback velocity do not tend to converge to the 2D values with increasing $L$. Instead, the 2D results show significant deviations, with the 3D results being approximately 50\% higher. This confirms the need for 3D simulations to evaluate the flashback phenomena in practical devices correctly and implies that 2D and 3D cases may exhibit different mechanisms of flashback.

\begin{figure}[!tbph]
    \centering
    \includegraphics[width=\textwidth]{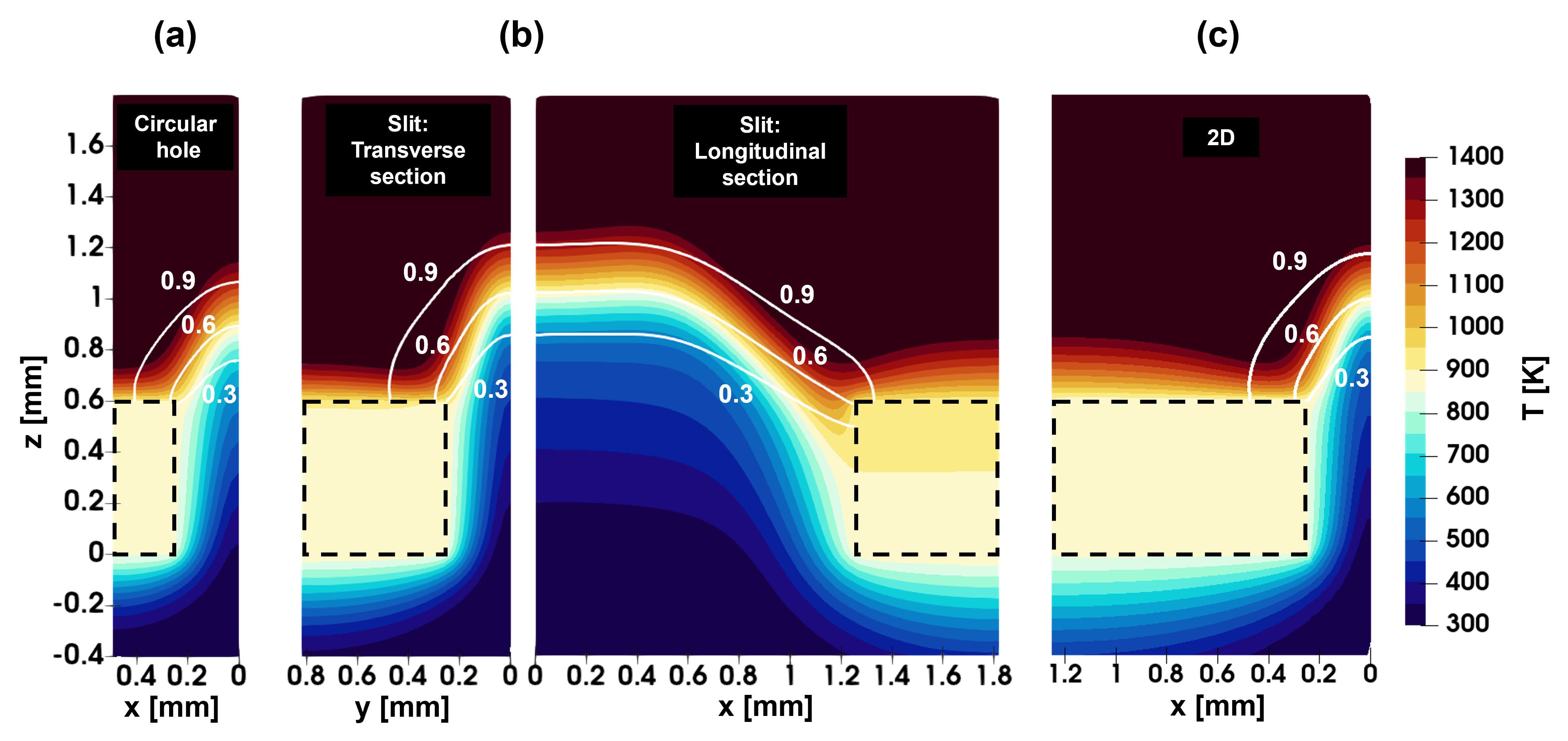}
    \caption{\footnotesize Temperature fields for (a) the circular hole, (b) the slit (${L=\SI{2}{\mm}}$), and (c) the infinite slit configurations at ${V_\mathrm{S}=\SI{3.3}{\m/\s}}$ and ${\phi=0.6}$. Progress variable iso-contours are plotted in white. The burner plate is delimited by a dashed line.}
    \label{fig:new_T}
\end{figure}

\begin{figure}[!tbph]
    \centering
    \includegraphics[width=\textwidth]{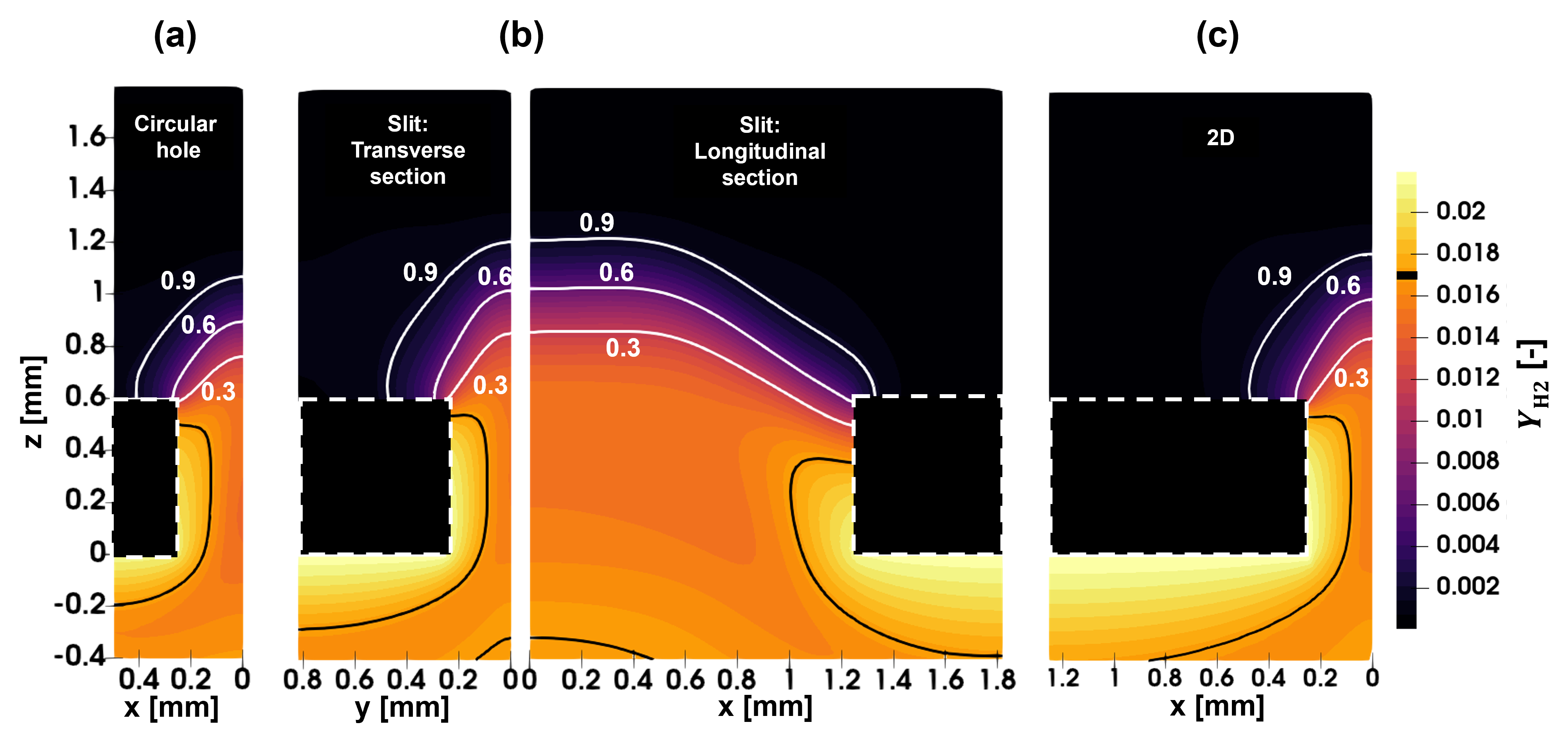}
    \caption{\footnotesize \ch{H2} mass fraction fields for (a) the circular hole, (b) the slit (${L=\SI{2}{\mm}}$), and (c) the infinite slit configurations at ${V_\mathrm{S}=\SI{3.3}{\m/\s}}$ and ${\phi=0.6}$. The nominal $Y_{\ch{H2}}$ is indicated by a black solid line. Progress variable iso-contours are plotted in white. The burner plate is delimited by a dashed line.}
    \label{fig:new_mf}
\end{figure}

\begin{figure}[!tbph]
    \centering
    \includegraphics[width=\textwidth]{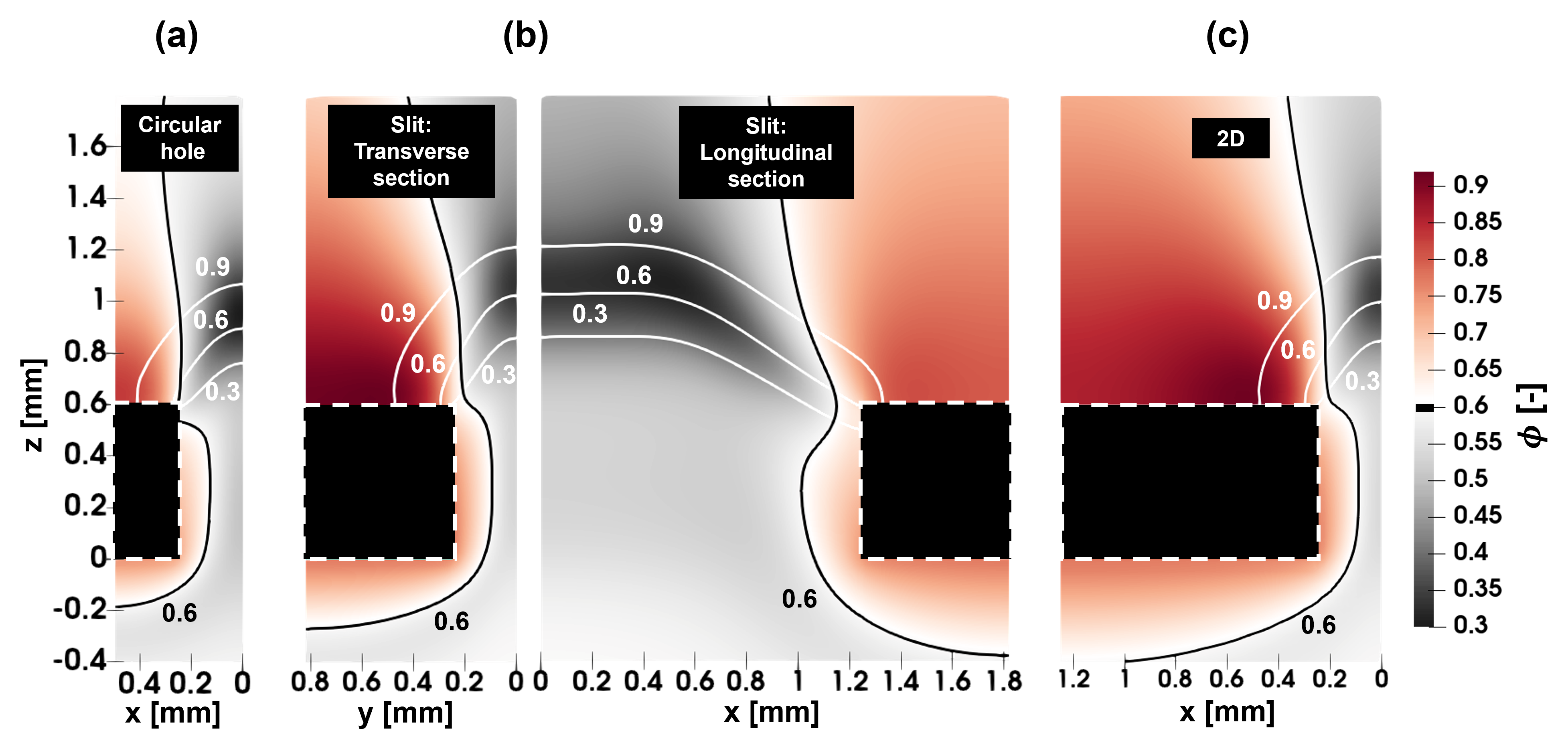}
    \caption{\footnotesize Local equivalence ratio fields for (a) the circular hole, (b) the slit (${L=\SI{2}{\mm}}$), and (c) the infinite slit configurations at ${V_\mathrm{S}=\SI{3.3}{\m/\s}}$ and ${\phi=0.6}$. The nominal $\phi$ is indicated by a black solid line. Progress variable iso-contours are plotted in white. The burner plate is delimited by a dashed line.}
    \label{fig:new_phi}
\end{figure}

\begin{figure}[!tbph]
    \centering
    \includegraphics[width=\textwidth]{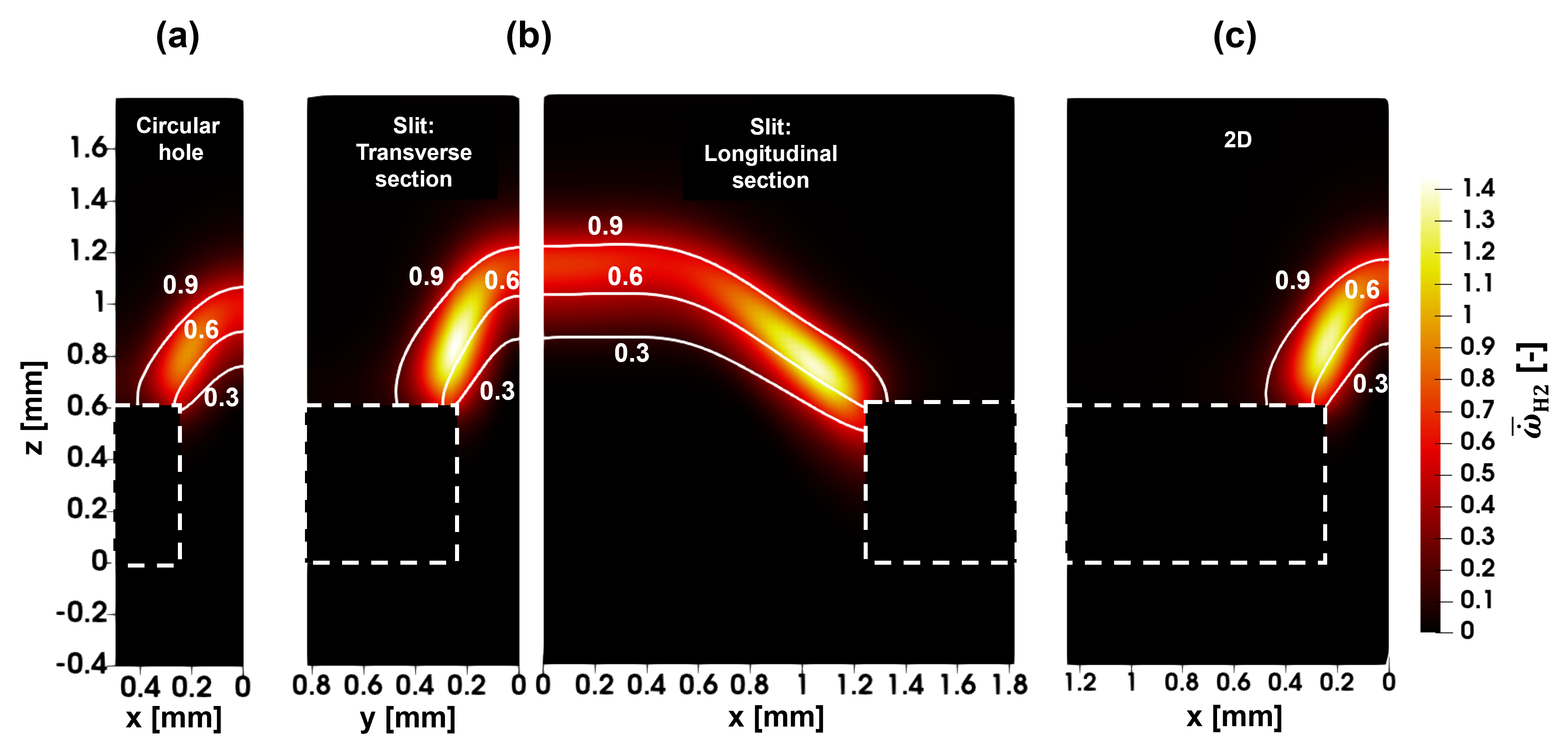}
    \caption{\footnotesize Normalized \ch{H2} consumption rate fields for (a) the circular hole, (b) the slit (${L=\SI{2}{\mm}}$), and (c) the infinite slit configurations at ${V_\mathrm{S}=\SI{3.3}{\m/\s}}$ and ${\phi=0.6}$. Progress variable iso-contours are plotted in white. The burner plate is delimited by a dashed line.}
    \label{fig:new_h2}
\end{figure}

\textcolor{black}{Hydrodynamic flashback in hydrogen flames within perforated burners is primarily governed by three key factors: preheating of the unburnt mixture, preferential diffusion due to non-unity Lewis numbers, and the Soret effect. The individual impact of these mechanisms on the propensity for flashback was studied separately and quantitatively analyzed in a previous work~\cite{FRUZZA_PROCI}.} Preheating, caused by heat transfer from the burner plate, increases both flow velocity and flame speed, with the latter rising more than linearly with temperature. This leads to a critical point where flame speed exceeds the flow velocity, triggering flashback. \textcolor{black}{Elevated wall and preheating temperatures not only enhance the flame speed but also significantly reduce the quenching distance, allowing flames to propagate through narrower slits. Unlike classical flashback theories such as the one by Lewis and von Elbe~\cite{LEWISVONELBE1943}, which typically assume cold walls, this behavior is peculiar to configurations with elevated wall temperatures, where ``quenching failure" flashback can become a dominant mechanism~\cite{PERS2024quenching}.} Preferential diffusion in low Lewis number mixtures enriches the flame base, increasing flame speed and promoting flashback~\cite{LAW19891381}. The Soret effect drives lighter species like \ch{H2} towards hotter regions, further enriching the mixture near the burner plate. This intensifies combustion at the flame base, leading to a self-accelerating feedback mechanism that promotes flashback~\cite{VANCEsoret,FRUZZA_PROCI}. The dominant effect among these varies with the specific geometry and operating conditions, substantially influencing the flame structure and the flashback mechanism. To clarify the distinct behaviors observed between the circular hole, the slit, and the infinite slit configurations, we compare their flame structures at a fixed inlet velocity, corresponding to the flashback limit of the slit. The flashback limit is defined as the lowest stable inlet velocity just before flashback occurs. In Figures~\ref{fig:new_T},~\ref{fig:new_mf},~\ref{fig:new_phi}, and~\ref{fig:new_h2}, we show the temperature, \textcolor{black}{\ch{H2} mass fraction,} local equivalence ratio, and normalized \ch{H2} consumption rate fields for a circular hole (a), a slit with a length of ${L=\SI{2}{\mm}}$ (b), and the infinite slit (c). For all cases, ${V_\mathrm{S}=\SI{3.3}{\m/\s}}$, corresponding to the flashback limit of the slit, and ${\phi=0.6}$. The local equivalence ratio is defined according to the Bilger formula~\cite{kee}. The molecular \ch{H2} consumption rate, $\dot{\omega}_{\ch{H2}}$, is normalized as $\Bar{\dot{\omega}}_{\ch{H2}} = \dot{\omega}_{\ch{H2}} / \mathrm{max}(\dot{\omega}_{\ch{H2},\mathrm{1D}})$, where $\mathrm{max}(\dot{\omega}_{\ch{H2},\mathrm{1D}}) = \SI{85.40}{\kg\per\cubic\metre\per\s}$ represents the peak \ch{H2} consumption rate for the corresponding unstretched 1D flame. To identify the flame front, iso-contours of the progress variable are included. The progress variable is defined as $C=1-Y_{\ch{H2}}/Y_{\ch{H2},u}$, where $Y_{\ch{H2}}$ denotes the mass fraction of \ch{H2} and $Y_{\ch{H2},u}$ its value in the unburnt mixture. 

\subsubsection*{Circular hole configuration}

In the circular hole configuration, preheating plays a dominant role in determining flame behavior. \textcolor{black}{As discussed in Section~\ref{sub:nr_flow}}, this configuration is the most confined of the three cases studied, characterized by the higher surface-to-volume ratio, $\sigma$. A higher surface-to-volume ratio enhances the preheating of unburnt gases entering the channel, which is reflected in an increased flame tip temperature. In the circular hole configuration, the temperature at the flame tip reaches approximately $T \simeq \SI{750}{K}$ at a progress variable of $C = 0.3$ (Figure~\ref{fig:new_T}(a)), about $\SI{150}{K}$ higher than in the slit configurations. Consequently, the flame tip is positioned closer to the hole exit. In contrast, preferential diffusion and the Soret effect have minimal impact for the circular hole configuration. The small channel size---the flame thickness is comparable with the hole radius---and high preheating produce a quasi-hemispherical flame with reduced curvature, which limits curvature-induced preferential diffusion effects. For the Soret effect, fuel enrichment is mainly concentrated near the channel inlet, where steep temperature gradients exist. However, further downstream, the gas temperature becomes more uniform, and the Soret-driven fuel migration toward the walls diminishes. As a consequence, no significant non-uniformities in the equivalence ratio or \ch{H2} consumption rate are observed along the flame front, indicating that preheating of the unburnt gases is the primary mechanism driving flashback in this case.

\subsubsection*{Slit configuration}

In the 3D slit configuration, preferential diffusion and the Soret effect play a significantly more prominent role. The extended flame front enhances curvature-induced preferential diffusion effects due to the differences in curvature between the flame tip and base. Additionally, along the long sides of the slit, the lower surface-to-volume ratio results in a lower unburnt gas temperature, creating pronounced temperature gradients toward the inner channel wall. These gradients amplify the influence of the Soret effect on flame morphology. Together with preferential diffusion and higher fuel availability compared to circular holes, substantial enrichment occurs at the flame base, \textcolor{black}{as demonstrated by the high \ch{H2} mass fraction close to the walls (Figures~\ref{fig:new_mf}(b))} and the equivalence ratio reaching up to $\phi=0.8$ along the walls and slit ends (Figures~\ref{fig:new_phi}(b)). This enrichment leads to a peak in the normalized \ch{H2} consumption rate, $\Bar{\dot{\omega}}_{\ch{H2}} \simeq 1.5$, at the flame base (Figures~\ref{fig:new_h2}(b)), increasing flame speed and promoting flashback. 

Notably, the slit ends emerge as the most critical regions for flashback initiation. In addition to the reduced velocity gradients caused by the slit’s aerodynamics, as detailed in Section~\ref{sub:nr_flow}, enrichment driven by preferential diffusion and the Soret effect, along with significant preheating, also plays a crucial role. Figure~\ref{fig:new_T}(b) illustrates a high-temperature region at the slit ends. As previously discussed, the surface-to-volume ratio at the slit ends is double that of the long sides, enhancing heat transfer and resulting in higher temperatures of the unburnt gases flowing through the slit ends relative to the central region. The lower velocity gradients, compared to the center of the slit, combined with the increased temperature at the slit ends, further elevate the flame speed, causing the flame to attach further upstream, closer to the inner surface of the channel. These combined effects of aerodynamics, preferential diffusion, the Soret effect, and enhanced preheating make the slit ends the most likely sites for flashback initiation in 3D slits. \textcolor{black}{We note that the spatial variability of preheating, even within a single configuration, complicates the definition of a representative preheating temperature. Therefore, our choice to define $V_\mathrm{S}$ and $s_L$ a priori based on nominal conditions ensures consistent comparisons across cases.}

\subsubsection*{Infinite slit configuration}

The 2D simulation accurately replicates the transverse section at the center of the 3D slit along the long side, showing almost identical flame front structures in both cases. However, the exclusion of the slit ends in the 2D configuration may result in a fundamentally different flashback dynamics compared to the finite-length slit, explaining the gap in flashback velocity.

\subsubsection*{Synthesis of key findings}

In summary, the 2D configuration, despite being significantly influenced by preferential diffusion and the Soret effect, experiences limited preheating due to the low surface-to-volume ratio, resulting in the lowest flashback velocity among the three cases. In the circular hole configuration, the confined geometry suppresses preferential diffusion and the Soret effect, but substantially enhances preheating of the unburnt gases, which plays a decisive role in increasing the flashback velocity compared to the 2D case. In the 3D slit configuration, both fuel enrichment and preheating are crucial factors. The elongated geometry promotes non-uniformities in the equivalence ratio and \ch{H2} consumption rate along the flame front. \textcolor{black}{The aerodynamics of the slit, particularly the lower velocity gradients at the slit ends, causes the flame to anchor further upstream compared to the long sides of the slit.} Additionally, the high surface-to-volume ratio at the slit ends amplifies preheating, making these regions the most likely points for flashback initiation, as both fuel enrichment and preheating increase the flame speed. Consequently, this results in a higher flashback velocity compared to the circular hole configuration. The dominant role of the slit ends also explains the weak dependence of flashback velocity on slit length, as flashback consistently originates at the slit ends, regardless of the total slit length.

\subsection{Flashback dynamics}\label{sub:dynamics}

The steady-state flame structures offer valuable insights into the mechanisms of flashback, as outlined in Section~\ref{sub:fb}, particularly by highlighting the differences between circular holes, slits, and infinite slits. These findings help explain the variations in flashback velocities shown in Figure~\ref{fig:Vfb_L}. However, a more detailed investigation of the actual flashback dynamics is required to fully account for the significant differences in flashback velocities between 2D and 3D configurations. To address this, two transient simulations are conducted: one for a 2D configuration and another for a 3D slit with a length of $L = \SI{2}{\mm}$. These simulations follow the methodology described in Section~\ref{sub:transient}. In both cases, the slit width is set to $W = \SI{0.5}{\mm}$, with a porosity of $\psi = 0.2$, an equivalence ratio of $\phi = 0.6$ is set for the inlet mixture. A detailed summary of the geometric and operating parameters is provided in Table~\ref{tab:dynamics}.
\begin{table}[htbp]
\centering
\caption{Overview of the simulation cases described in Section~\ref{sub:dynamics}.}
\label{tab:dynamics}
{\small\begin{tabular}{@{}ccccccc@{}}\toprule
\multicolumn{5}{c}{\textbf{Geometry}}&\multicolumn{1}{c}{\textbf{Op. parameters}}\\ \midrule
\textbf{Config.}&$\boldsymbol{L}$ [mm]&$\boldsymbol{W}$ [mm]&$\boldsymbol{\psi}$&\textbf{$\boldsymbol{H}$}&$\boldsymbol{\phi}$\\ \midrule
2D&-&0.5 mm&0.2&0.6 mm&0.6\\
\multirow{2}{*}{3D}&\multirow{2}{*}{\shortstack{2 mm}}&\multirow{2}{*}{0.5 mm}&\multirow{2}{*}{0.2}&\multirow{2}{*}{0.6 mm}&\multirow{2}{*}{0.6}\\
&&&&&&\\ \bottomrule
\end{tabular}}
\end{table}

\begin{figure}[!tb]
    \centering
    \includegraphics[width=0.9\textwidth]{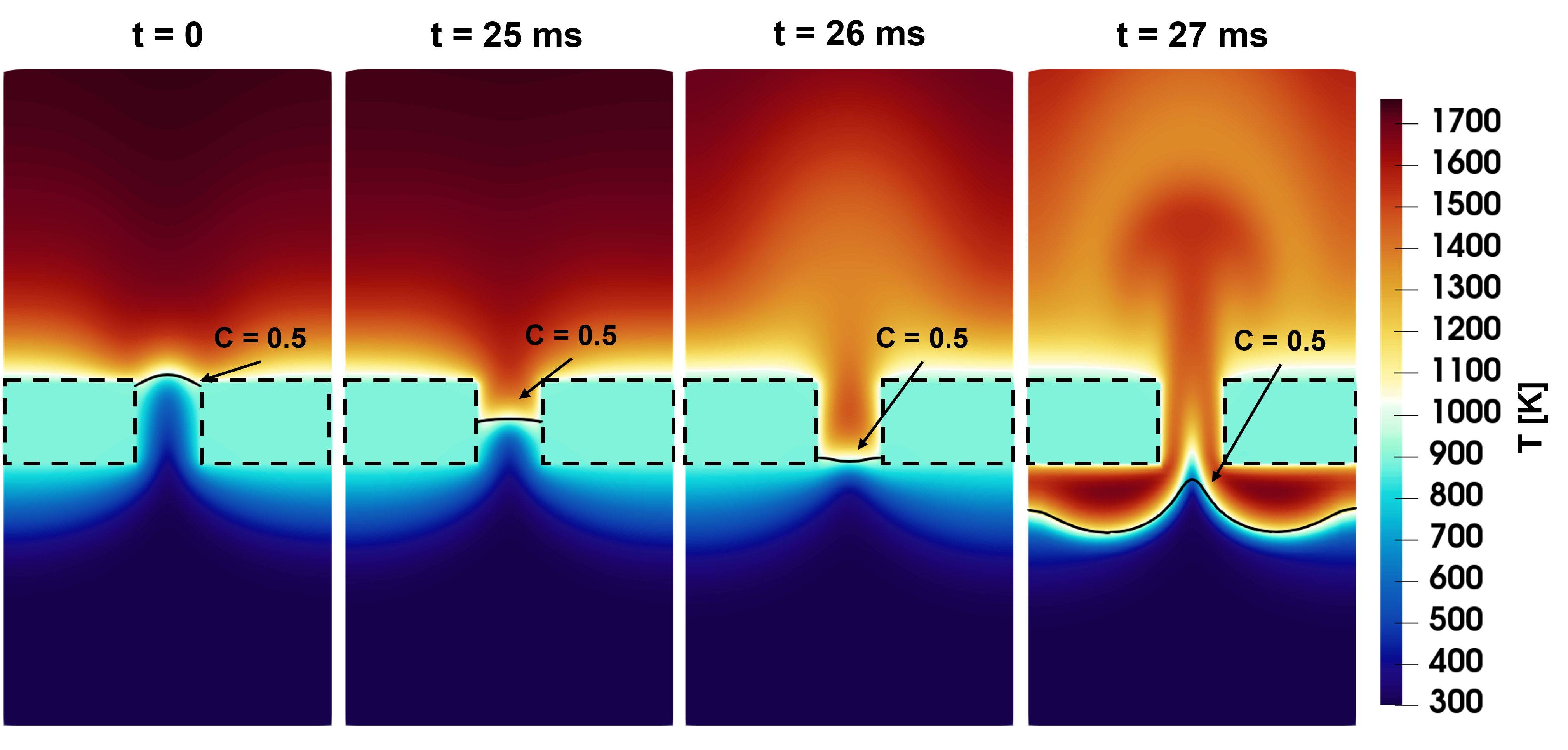}
    \caption{\footnotesize Evolution of the temperature profile at four distinct time points during the onset of flashback in the 2D configuration. Progress variable iso-contours corresponding to $C=0.5$ are plotted in black. The burner plate is delimited by a dashed line.}
    \label{fig:snap_2D}
\end{figure}

In Figure~\ref{fig:snap_2D}, we illustrate the flashback event for the 2D configuration. The evolution of the temperature fields is displayed in four snapshots taken during the occurrence of flashback. An iso-contour of progress variable corresponding to $C=0.5$ is included to identify the flame front.
The initial instant, marked as $t=0$, corresponds to when $V_\mathrm{S}$ is decreased from the last stable velocity to the flashback velocity $V_\mathrm{FB}=\SI{2.0}{\m/\s}$. At this moment, the flame front is positioned very close to the slit exit. After $\SI{25}{\ms}$, the flame front becomes flat and has moved upstream to the middle of the channel. Within the next $\SI{2}{\ms}$, the flame front crosses the entire channel symmetrically. In these simulations, the dynamics can vary from symmetrical to asymmetrical, depending on the operating and geometric conditions, as highlighted in previous studies~\cite{FRUZZA2023,fruzzaUQ,FLORESMONTOYA2023113055}. Nevertheless, 2D simulations consistently predict that flashback initiates along the long side of the slit. However, in 3D slits, the presence of boundaries at their ends may induce significantly different flashback dynamics, which the 2D simulations cannot capture.

\begin{figure}[!tb]
    \centering
    \subfigure[]{\includegraphics[width=0.45\textwidth]{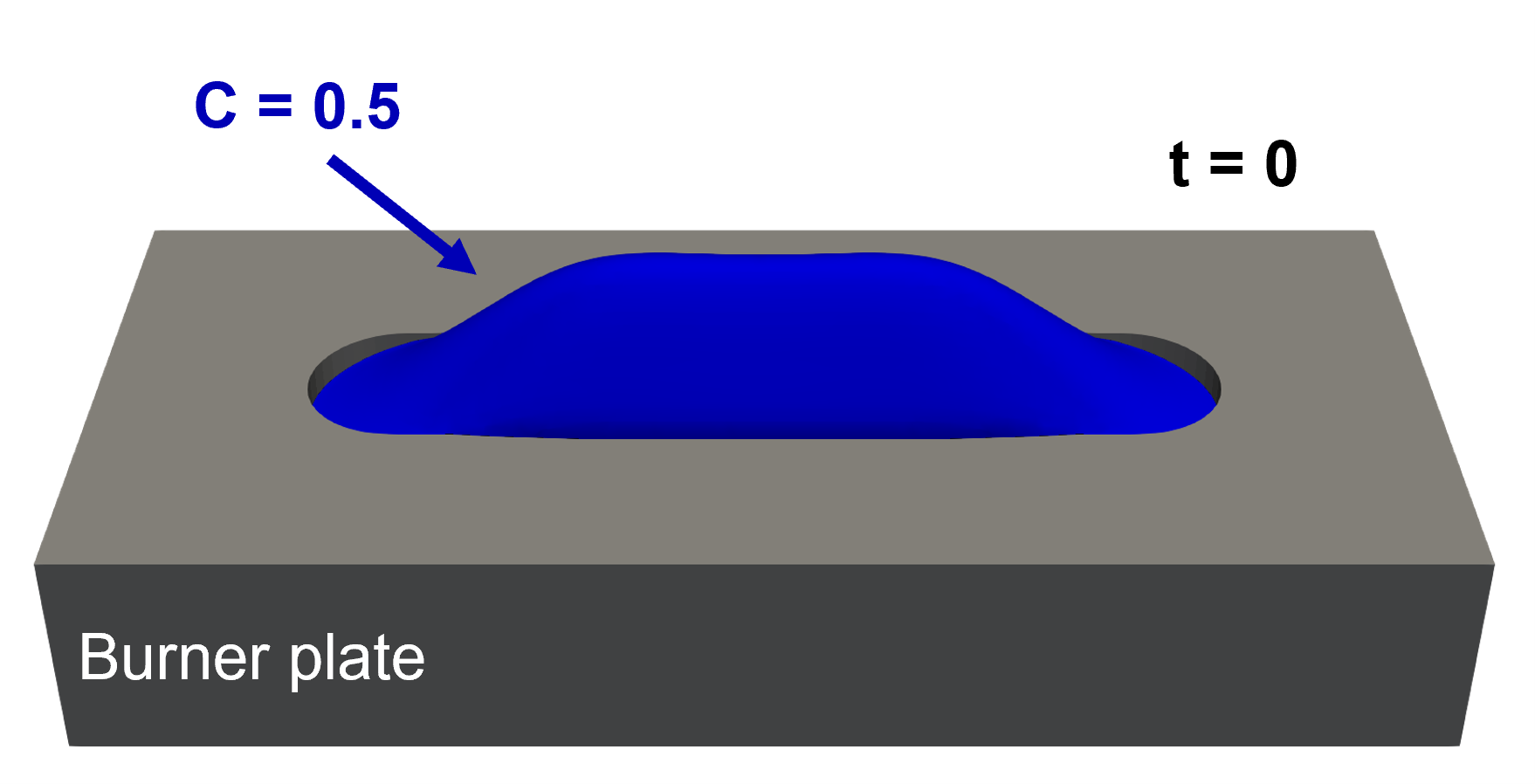}}
    \subfigure[]{\includegraphics[width=0.45\textwidth]{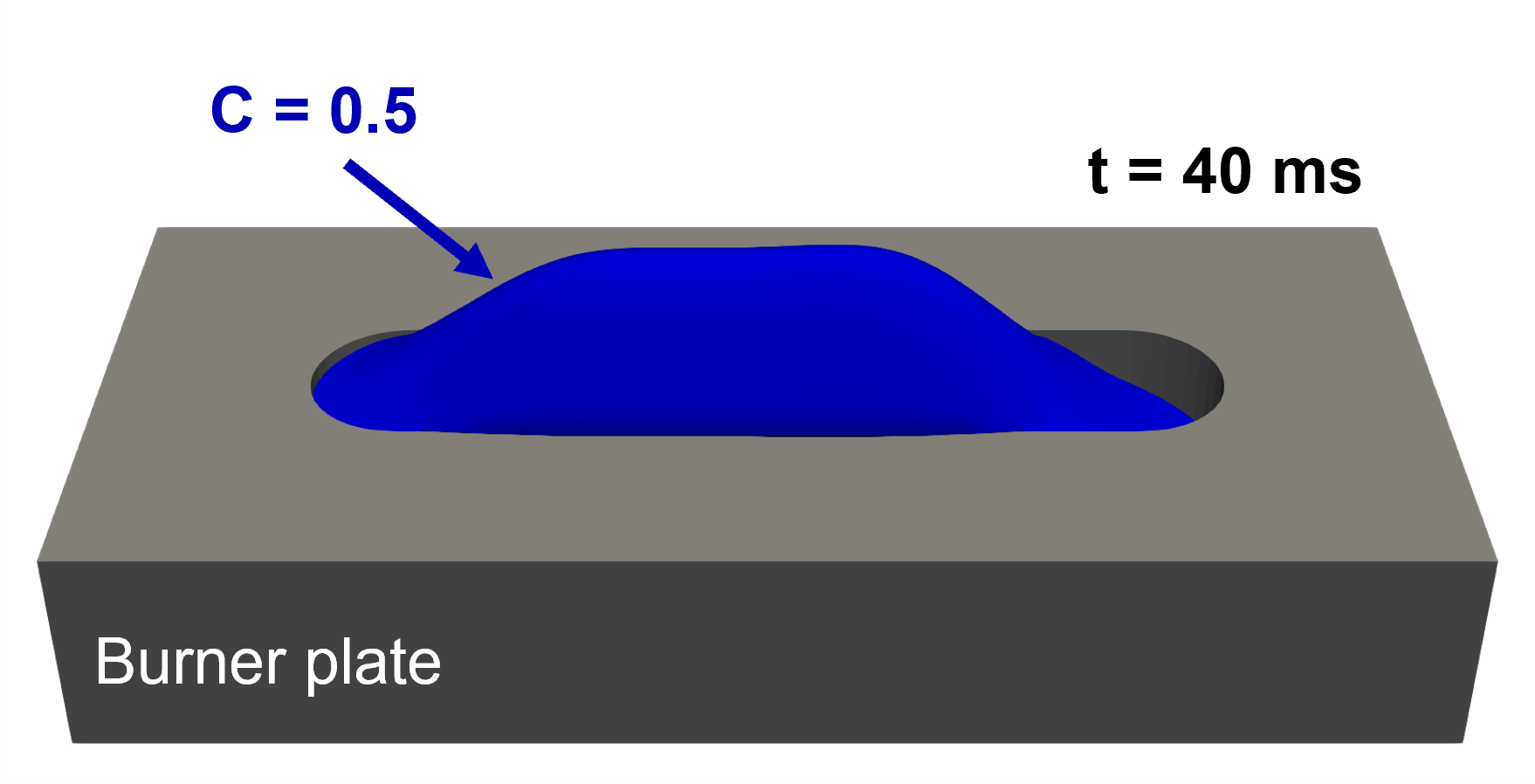}}
    \subfigure[]{\includegraphics[width=0.45\textwidth]{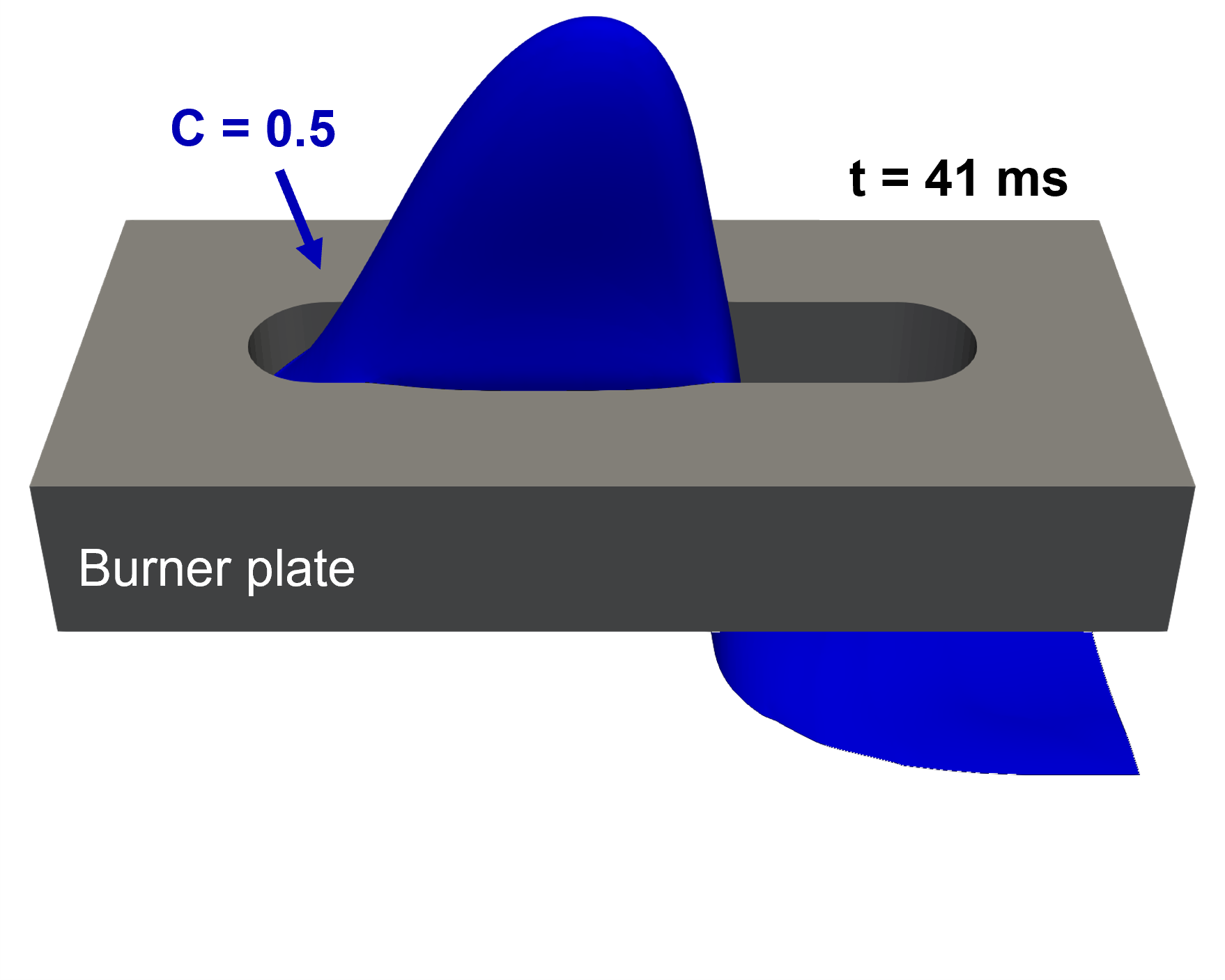}}
    \subfigure[]{\includegraphics[width=0.45\textwidth]{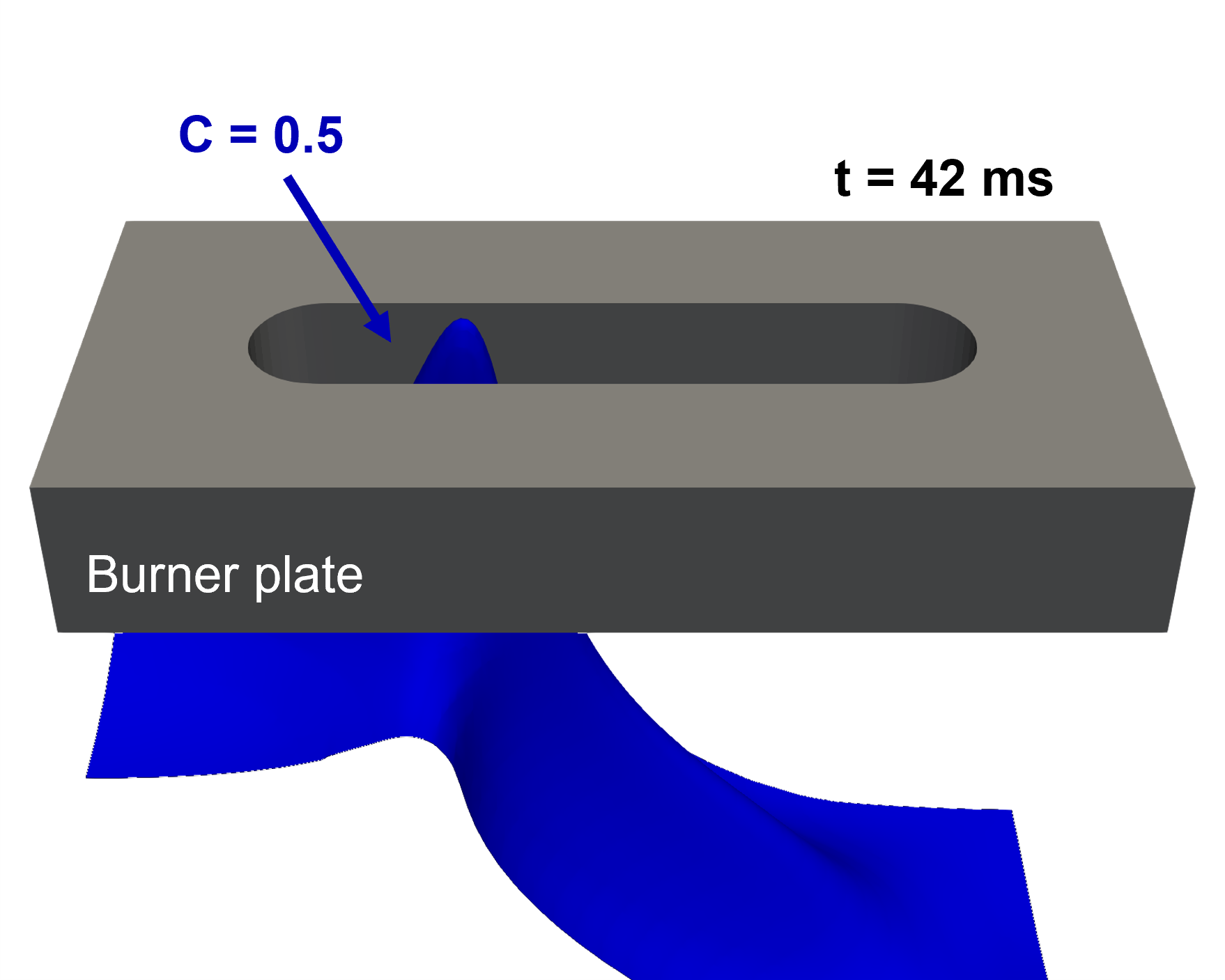}}
    \caption{\footnotesize Flame front evolution at four distinct times during the onset of flashback in the three-dimensional slit with $L=\SI{2}{\mm}$. Iso-surfaces of progress variable at $C=0.5$ are shown to visualize the flame front.}
    \label{fig:slit_FB}
\end{figure}

Figure~\ref{fig:slit_FB} depicts a sequence of four snapshots captured during the flashback occurrence for the 3D configuration. The snapshots illustrate the evolution of the flame front, defined as an iso-surface of progress variable corresponding to $C=0.5$.
The initial instant, denoted as $t=0$, marks the moment of the reduction of the inlet velocity from the last stable velocity to the flashback velocity $V_\mathrm{FB}=\SI{3.3}{\m/\s}$. At $t=0$, the flame front is situated at its initial stable position (Figure~\ref{fig:slit_FB}(a)). During the first ${\SI{40}{ms}}$, there are minimal changes in the position of the flame front. By ${t =\SI{40}{ms}}$, flashback initiates rapidly within the slit. The initiation is asymmetric, starting at one end of the slit (Figure~\ref{fig:slit_FB}(b)). From there, the flame front moves backward towards the inlet, traversing the slit entirely on that side (Figure~\ref{fig:slit_FB}(c)), and ultimately driving the entire flashback process (Figure~\ref{fig:slit_FB}(d)). The asymmetrical dynamics and the initiation of the flashback at one of the slit ends are consistent with the experimental observations of Pers et al.~\cite{PERS2024autoignition}, where the same initiation point for the ``hydrodynamic" flashback is observed.

\begin{figure}[!tb]
    \centering
    \subfigure[]{\includegraphics[width=0.45\textwidth]{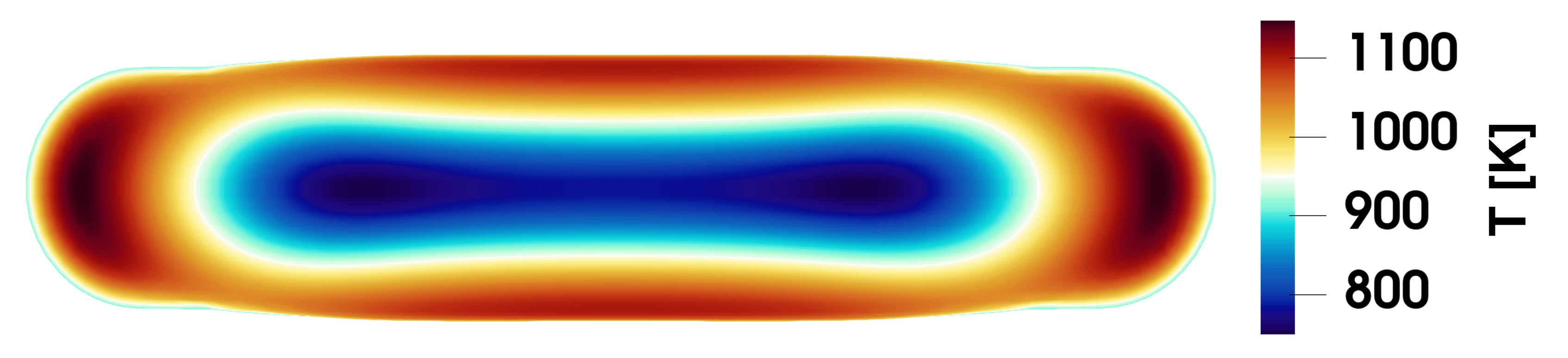}}
    \subfigure[]{\includegraphics[width=0.45\textwidth]{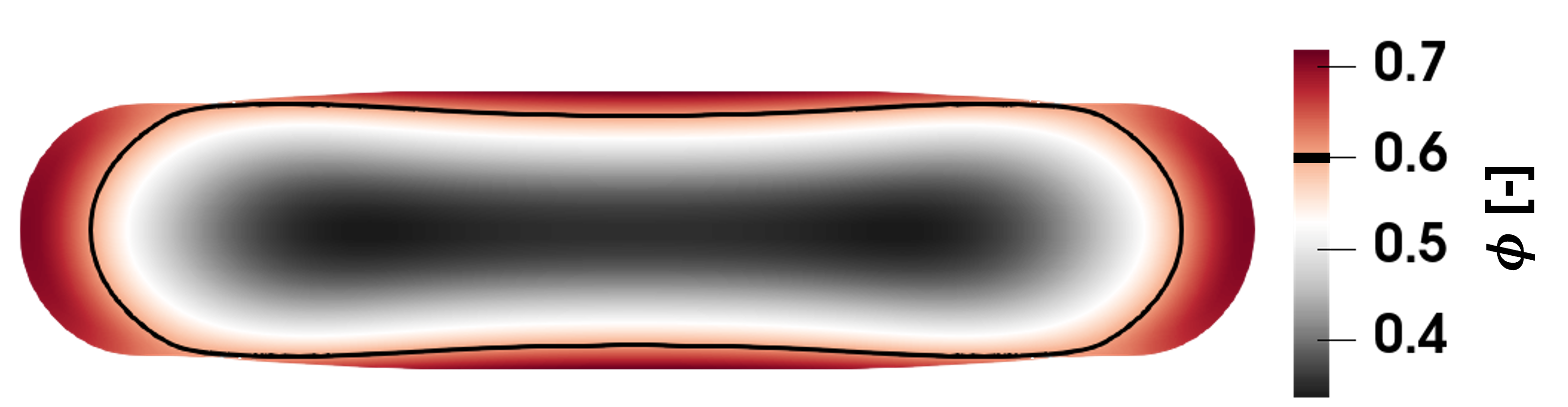}}
    \subfigure[]{\includegraphics[width=0.45\textwidth]{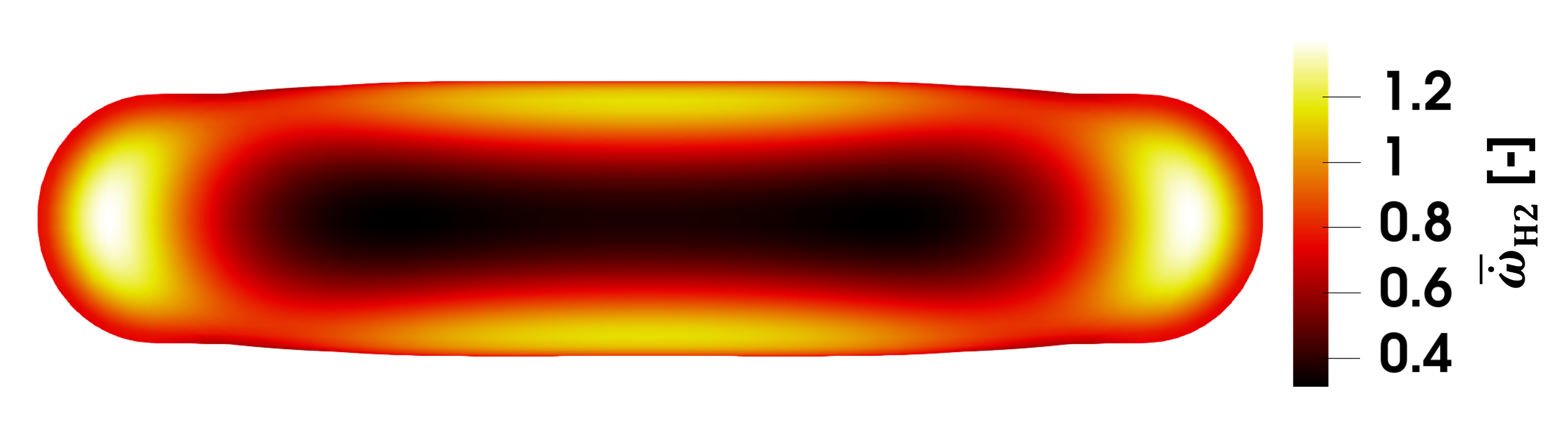}}
    \subfigure[]{\includegraphics[width=0.45\textwidth]{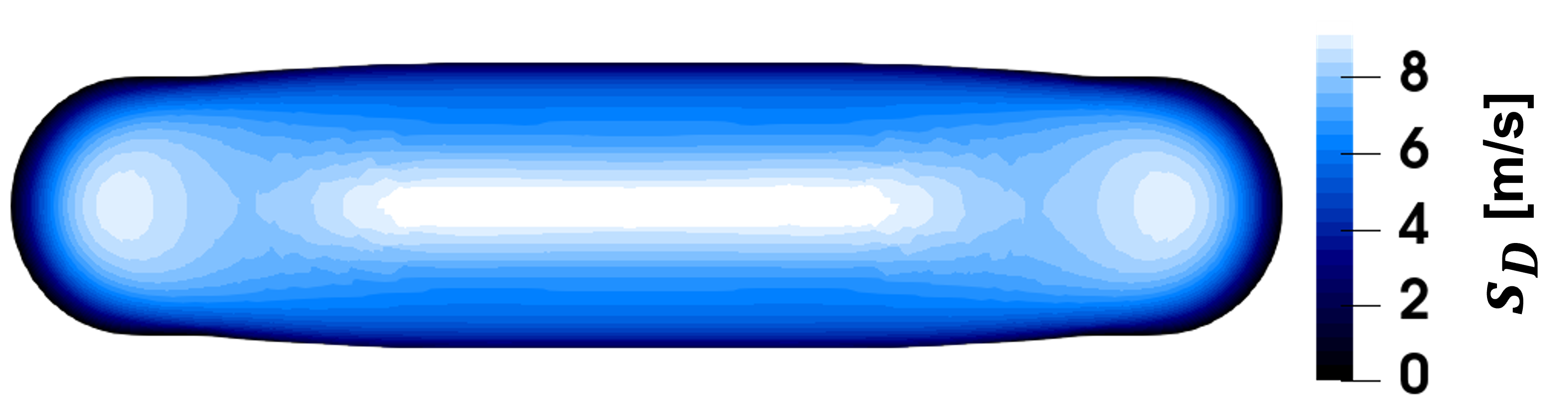}}
    \caption{\footnotesize Iso-contour of progress variable $C=0.5$ for a slit with ${L = \SI{2}{\mm}}$ at the flashback limit for the case ${\phi=0.6}$ (top view). The iso-contours are colored by: (a) temperature, (b) local equivalence ratio, (c) normalized \ch{H2} consumption rate, and (d) displacement speed.}
    \label{fig:slit_fields}
\end{figure}

For better visualization of the critical physical mechanisms involved, a colored image of the last stable flame, corresponding to Figure~\ref{fig:slit_FB}(a), is presented in Figure~\ref{fig:slit_fields}. The iso-contour of the progress variable at $C=0.5$ is colored to show temperature (a), local equivalence ratio (b), normalized molecular \ch{H2} consumption rate (c), and displacement speed (d), defined as $s_\mathrm{D}=\boldsymbol{v}\cdot\boldsymbol{n}$, where $\boldsymbol{n}$ is the unit vector normal to the iso-surface.
A temperature peak is observed at the slit ends where the higher surface-to-volume ratio results in greater preheating of the mixture (Figure~\ref{fig:slit_fields}(a)). In these regions, preferential diffusion and the Soret effect contribute to fuel enrichment of the mixture (Figure~\ref{fig:slit_fields}(b)). These phenomena lead to an increase in the \ch{H2} consumption rate (Figure~\ref{fig:slit_fields}(c)), and subsequently, an increase in flame speed, as indicated by the local maxima of displacement speed (Figure~\ref{fig:slit_fields}(d)). \textcolor{black}{Consequently, this region becomes the initiation zone for flashback, as the combination of greater preheating, fuel enrichment, and lower velocity gradients at the slit ends allows the local flame speed to surpass the flow velocity, triggering flashback initiation.}

The transient simulations confirm the critical role of the slit ends in flashback initiation, reinforcing the explanation for the weak dependence of flashback velocity on slit length beyond ${L=\SI{1}{\mm}}$, as shown in Figure~\ref{fig:Vfb_L}. With flashback consistently starting at the slit ends, the specific length, $L$, becomes secondary. The results also highlight the inherently three-dimensional nature of the flashback process, which cannot be fully captured by 2D configurations. 

\section{Conclusions}

In this study, both steady-state and transient 3D simulations are conducted to examine the flashback velocities and dynamics within a single slit in a hydrogen-fueled premixed perforated burner. \textcolor{black}{Non-reactive simulations are performed to investigate the aerodynamics and heat transfer mechanisms in the circular hole and slit configurations.} Steady-state simulations are then utilized to assess the impact of varying slit lengths on flame shape and burner plate temperature, with the results being compared with those from 2D configurations to evaluate the validity of the infinite slit approximation. Additionally, flashback velocities are computed at three different equivalence ratios using a steady-state approach, and these results are compared with predictions from 2D simulations. Detailed transient simulations are also employed to explore the temporal evolution of flashback dynamics in a three-dimensional slit.

\textcolor{black}{The non-reactive analysis highlights distinct aerodynamic and thermal behaviors between the circular hole and slit configurations. The circular hole exhibits higher velocities and steeper velocity gradients, with its high surface-to-volume ratio significantly enhancing heat transfer efficiency and preheating. In contrast, the slit configuration exhibits pronounced internal variations. The central region is characterized by relatively low temperatures and elevated velocity gradients in the transverse direction, while the slit ends experience intensified heating and reduced velocity gradients, ultimately producing conditions comparable to those observed for the circular hole. These differences underscore the importance of considering both global and localized aerodynamic and thermal effects when analyzing such configurations.}

Additionally, it is found that while 2D simulations can accurately predict flame front structures and burner plate temperature for long slits, they fail to capture the thermal effects caused by the slit ends in shorter configurations. Even though the results of 2D simulations converge with the 3D results as the slit length approaches infinity, it becomes less reliable for slits of practical length due to the pronounced influence of the slit ends.

The results for flashback velocities indicate that the circular hole geometry demonstrates greater resistance to flashback. As the slit length increases beyond ${L=\SI{1}{\mm}}$, the flashback velocity shows only a weak dependence on slit length. These findings are compared with those from a 2D configuration, revealing that the 3D results do not converge with the 2D predictions at large slit lengths. Notably, 2D simulations significantly underestimate the flashback velocity. While the infinite slit configuration is influenced by preferential diffusion and the Soret effect, its lower surface-to-volume ratio limits preheating, resulting in the lowest flashback velocities. In contrast, the circular hole geometry exhibits higher flashback velocities due to enhanced preheating, despite its confinement suppressing preferential diffusion and the Soret effect. The 3D slit configuration yields the highest flashback velocities, with the regions near the slit ends playing a crucial role due to combined effects of fuel enrichment, preheating, \textcolor{black}{and favorable aerodynamics}. This highlights why the asymptotic limit of flashback velocity for large slit lengths differs from that of an infinite slit in the 2D configuration, where the influence of the slit ends is consistently neglected. Moreover, the dominant influence of the slit ends also explains the weak dependence of the flashback velocity on the slit length, as the critical region for flashback initiation remains at the ends, regardless of the overall slit length.

Finally, the transient simulations validate the central role of the slit ends in flashback initiation and reveal an asymmetric flashback dynamics, where flashback begins at one end of the slit. In contrast, the 2D simulations predict flashback initiation along the long sides of the slit, missing the three-dimensional effects introduced by the boundaries at the slit ends. These transient results further underscore the inherently three-dimensional nature of the process, which cannot be captured by 2D simulations.

\section*{Author contributions}

Filippo Fruzza: Conceptualization, Methodology, Data analysis, Writing. Hongchao Chu: Conceptualization, Reviewing and editing. Rachele Lamioni: Conceptualization, Methodology, Reviewing and editing. Temistocle Grenga: Reviewing and editing. Chiara Galletti: Conceptualization, Reviewing and editing. Heinz Pitsch: Conceptualization, Reviewing and editing.

\section*{Acknowledgements}

This research is partially funded by the Ministry of University and Research (MUR) and Immergas S.p.A., Brescello, RE (Italy), as part of the PON 2014-2020  ``Research and Innovation" resources - Green/Innovation Action - DM MUR 1061/2021 and DM MUR 1062/2021.

\bibliography{FB_3D}

\end{document}